\documentclass[journal, 12pt, draftclsnofoot, onecolumn]{IEEEtran}

\usepackage[T1]{fontenc}
\usepackage{hyperref}
\usepackage{multirow}
\usepackage{diagbox}
\usepackage{cite}
\usepackage[pdftex]{graphicx}
\usepackage{graphicx, subfigure}
\usepackage{amsmath}
\usepackage{float}
\usepackage{subfloat}
\usepackage{booktabs}
\usepackage{stfloats}
\usepackage{footnote}
\usepackage{threeparttable}
\usepackage{color}

\begin{document}

\title{Model-Driven DNN Decoder for Turbo Codes: Design, Simulation and Experimental Results}

\author{\normalsize {Yunfeng~He,~Jing~Zhang,~Shi~Jin, Chao-Kai~Wen, and Geoffrey Ye Li}

\thanks{Parts of this work were accepted for presentation in the 16th IEEE Asia Pacific Wireless Communications Symposium 2019 (APWCS 2019), Singapore University of Technology and Design, Singapore.}
\thanks{Y.~He, J.~Zhang and S.~Jin are with the National Mobile Communications Research
Laboratory, Southeast University, Nanjing, 210096, P. R. China (email: heyunfeng@seu.edu.cn, jingzhang@seu.edu.cn, jinshi@seu.edu.cn).}
\thanks{C.-K.~Wen is with the Institute of Communications Engineering, National Sun Yat-sen University, Kaohsiung 80424, Taiwan (e-mail: chaokai.wen@mail.nsysu.edu.tw).}
\thanks{G.~Y.~Li is with the School of Electrical and Computer Engineering, Georgia Institute of Technology, Atlanta, GA 30332 USA (e-mail: liye@ece.gatech.edu).}
}

\maketitle
\begin{abstract}
	This paper presents a novel model-driven deep learning (DL) architecture, called TurboNet, for turbo decoding that integrates DL into the traditional max-log-maximum \emph{a posteriori} (MAP) algorithm. The TurboNet inherits the superiority of the max-log-MAP algorithm and DL tools and thus presents excellent error-correction capability with low training cost. To design the TurboNet, the original iterative structure is unfolded as deep neural network (DNN) decoding units, where trainable weights are introduced to the max-log-MAP algorithm and optimized through supervised learning. To efficiently train the TurboNet, a loss function is carefully designed to prevent tricky gradient vanishing issue. To further reduce the computational complexity and training cost of the TurboNet, we can prune it into TurboNet+. Compared with the existing black-box DL approaches, the TurboNet+ has considerable advantage in computational complexity and is conducive to significantly reducing the decoding overhead. Furthermore, we also present a simple training strategy to address the overfitting issue, which enable efficient training of the proposed TurboNet+. Simulation results demonstrate TurboNet+’s superiority in error-correction ability, signal-to-noise ratio generalization, and computational overhead. In addition, an experimental system is established for an over-the-air (OTA) test with the help of a 5G rapid prototyping system and demonstrates TurboNet’s strong learning ability and great robustness to various scenarios.
\end{abstract}


%
\IEEEpeerreviewmaketitle

\section{Introduction}
{\label{section: Introduction}
\linespread{1.55}\selectfont
To address channel distortion and interference, error correction codes (ECC) have been widely used in communication systems to improve the reliability of data transmission.
The overview in~\cite{channel coding: Costello} provides detail evolution of channel coding from Hamming codes to capacity-approaching codes.
The algebraic coding paradigm dominated the field of channel coding for the first few decades.
An infinite class of single-error-correcting binary linear codes was proposed in~\cite{Error:Hamming}. In 1954, Reed--Muller codes~\cite{Application: Muller} were developed.
Later on, some related codes, such as the Bose--Chaudhuri--Hocquenghem and Reed--Solomon codes
\cite{On: Bose, Polynormial: Reed} were invented.
Another branch is probabilistic coding.
A classic scheme in this field, convolutional codes, was introduced by Peter Elias in 1955~\cite{Coding: Peter}.
Two capacity-approaching codes were found a while ago: turbo codes~\cite{IEEEturbocode:Berrou} and low-density parity-check codes~\cite{LDPC: Gallager}. Recently, polar codes, the first class of capacity-achieving codes, were invented by Erdal Arikan~\cite{Polar: Arikan}. Although turbo codes have been successfully applied in 3G and 4G systems due to their high reliability, its latency is too large to meet the low latency requirements of 5G systems. Reducing latency of decoding has always been a tricky task.

Recently, deep learning (DL) has made remarkable achievements in computer vision and natural language processing. Artificial intelligence has been considered one of the key technologies for the next generation of mobile communication systems and can learn potential models in a data-driven method, thus avoiding an imprecise hypothesis.
The intelligent communication community has attained numerous accomplishments~\cite{Physical: Shea, DL for wireless physical layer: Wang, DL in physical layer: Qin}, including channel estimation (CE)~\cite{Over-the-air: Shea}, signal detection (SD)~\cite{MIMO detection: He}, CE and SD combined orthogonal frequency division multiplexing (OFDM) receivers~\cite{Power: Ye, ComNet: Gao, AI receiver: Zhang}, channel state information feedback~\cite{CSINet: Wen, CSINet: Wang}, autoencoder-based end-to-end communication systems~\cite{DL-based communication: Dorner, Channel: Ye}, and channel coding\cite{IEEEondeep:Gruber, IEEEanartificial:Wang, IEEEcommunicationalgorithms:Kim, Denoise: Cao, IEEElearningto:Nachmani, IEEEneuraloffset:Lugosch, IEEEdeeplearning:Nachmani, Active learning: Beery, He-Channel coding}. 

In general, DL-based channel decoding is in the preliminary exploration stage. The data-driven DL approach in~\cite{IEEEondeep:Gruber} converts the decoding task into the pure idea of \emph{learning to decode} by optimizing the general black-box fully connected deep neural network (FC-DNN). Despite the advantage of one-shot decoding (i.e., no iterations), the FC-DNN based decoder is unable to exploit expert knowledge, a feature, which, in turn, renders the FC-DNN decoder unaccountable, and is fundamentally restricted by its dimensionality. Training any neural network in practice is extremely difficult because the training complexity increases exponentially along with the block length (e.g., for a turbo code with length of $K=40$, there are ${{2}^{40}}$ different codewords)~\cite{IEEEanartificial:Wang}. 
In~\cite{IEEEcommunicationalgorithms:Kim}, a recurrent neural network (RNN) architecture, containing two layers of bidirectional gated recurrent units, is adopted to learn the Bahl--Cocke--Jelinek--Ravi (BCJR) algorithm. A residual neural network decoder is designed in~\cite{Denoise: Cao} for polar codes where a denoising module based on residual learning is introduced. The proposed residual learning denoiser can remove a remarkable amount of noise from the received signals. The aforementioned data-driven decoding methods depend on a large amount of data to train numerous parameters, thereby converging slowly and suffering from high computational complexity.

To address the aforementioned issues, the model-driven DL approach can be used instead. Model-driven DL~\cite{DL in physical layer: Qin} is particularly suitable to iterative approaches in signal precessing and communications.
The concept of a ``soft'' Tanner graph has been introduced in~\cite{IEEElearningto:Nachmani}, in which weights are assigned to the Tanner graph of the belief propagation (BP) algorithm to obtain a partially connected DNN, that is, weighted BP (WBP) decoding. These weights are learned to facilitate proper weight message transmission in the Tanner graph, thereby improving the performance of the BP algorithm. To reduce the number of multiplications in~\cite{IEEElearningto:Nachmani}, a min-sum algorithm with trainable offset parameters has been proposed in~\cite{IEEEneuraloffset:Lugosch}. The aforementioned DNN-based BP decoder has been transformed into a RNN architecture in~\cite{IEEEdeeplearning:Nachmani}, named the BP-RNN decoder, by unifying the weights in each iteration and consequently reducing the number of parameters without sacrificing performance, where a trainable relaxation factor has been introduced to improve the performance of this BP-RNN decoder. In~\cite{Active learning: Beery}, active learning has been first applied in the ECC field to improve the WBP decoding.

In summary, there are two inherent limitations in the existing DL-based decoding methods. First, current data-driven approaches rely on vast training parameters and suffer extremely high computational complexity. Second, it is unknown whether the existing model-driven algorithms could be applied to sequential codes (e.g., turbo codes) for performance improvement, as the aforementioned model-driven decoding methods are all based on BP. Designing a model-driven decoding algorithm for turbo codes is of great significance for low-latency communications. To address these limitations, this paper introduces a novel model-driven DL architecture, called TurboNet, for turbo decoding, which integrates DL into the traditional max-log-maximum \emph{a posteriori} (MAP) algorithm. TurboNet is constructed according to the domain knowledge in traditional turbo decoding and employs model-driven DL to address the inherent limitations of the existing methods. Our contributions are listed as follows:
\begin{enumerate}
	\item[$\bullet$] The original iterative architecture is unfolded to obtain an ``unrolled'' (i.e., each iteration is considered separately) structure and the max-log-MAP algorithm is parameterized. Given such structure and a well-designed loss function, the parameters can be optimized via training data more efficiently than with the black-box FC-DNN~\cite{IEEEanartificial:Wang} approach. The proposed network works well for codes with large block length, whereas the FC-DNN fails to work.
	\item[$\bullet$] The TurboNet is then pruned into TurboNet+ to reduce computational complexity and further improve the error-correction capability. The TurboNet+ exhibits better performance compared with the max-log-MAP algorithm for turbo decoding with different code lengths, code rates, and modulation modes. It also contains considerably fewer parameters compared with the data-driven neural BCJR decoder proposed in~\cite{IEEEcommunicationalgorithms:Kim}.
	\item[$\bullet$] The overfitting issue of the TurboNet+ is studied in detail and a simple and effective training strategy is developed to train the TurboNet+ better and faster. The trained TurboNet+ decoder shows strong generalization to the signal-to-noise ratio (SNR).
	\item[$\bullet$] An experimental system for an over-the-air (OTA) test is established and extensive experimental results have demonstrated TurboNet+'s great flexibility and robustness, which facilitate its practical use.
\end{enumerate}

The rest of this paper is organized as follows. To obtain the model-driven DL architecture for turbo decoding, we briefly describe the system model in Section~\ref{section:system model}, including the turbo encoder structure, channel, and traditional turbo decoder based on the max-log-MAP algorithm. The architecture of the TurboNet is explained in Section~\ref{section: TurboNet Architecture}, including a redefined function that evaluates network loss. We investigate network pruning and present detailed training strategy in Section~\ref{TurboNet+ design}. The performance of the TurboNet in simulation and experimental scenarios is demonstrated in Sections~\ref{Simulation results and Discussion} and~\ref{OTA Test}, respectively. Section~\ref{Conslusion} concludes the contributions.

Notations: Column vectors are denoted by boldface letters. For a vector ${\mathbf{x}}$ of length $K$, $F(\mathbf{x})$ denotes ${(F({{x}_{1}}),~F({{x}_{2}}),~\ldots,~F({{x}_{K}}))}$, and ${F(\mathbf{x} |\mathbf{y})}$ denotes ${(F({{x}_{1}}|\mathbf{y}),~F({{x}_{2}}|\mathbf{y}),~\ldots,~F( {{x}_{K}}|\mathbf{y}))}$.}
\section{System Model}
\label{section:system model}
For the convenience of introducing the TurboNet architecture, we briefly describe turbo encoder, channel, and turbo decoder in this section.
\vspace{-0.2cm}
\subsection{Turbo Encoder}
\begin{figure}[tb]
	\centering
	\includegraphics[width=6in]{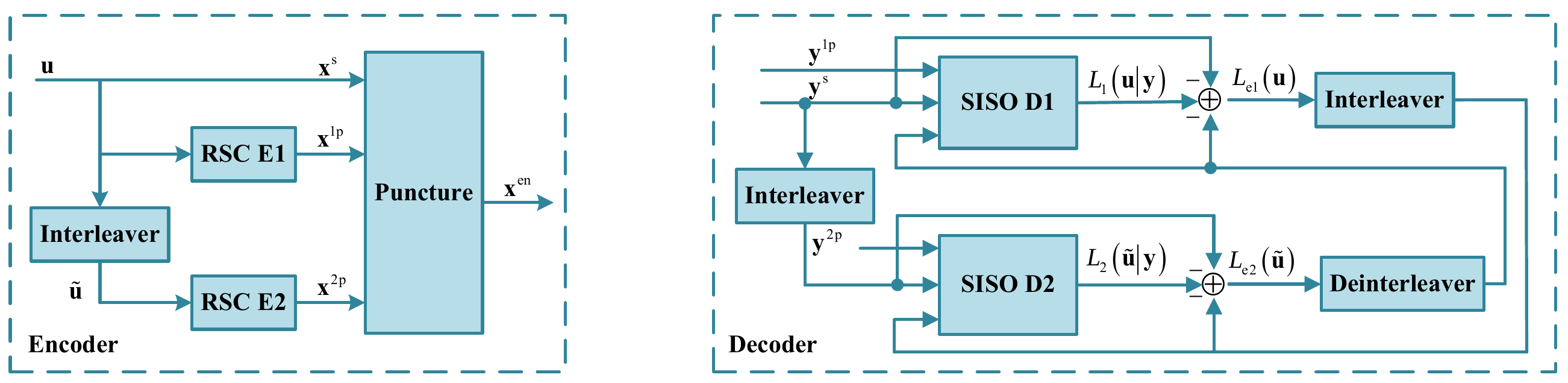}
	\caption{The structures of turbo encoder and decoder.}
	\label{Fig: turbo_encoder_decoder}
\end{figure}
Denote ${\mathbf{u}=({{u}_{1}},{{u}_{2}},\ldots,{{u}_{K}})}$ as the binary information sequence at the transmitter, which is interleaved into ${\tilde{\mathbf{u}}= ({{{\tilde{u}}}_{1}},{{{\tilde{u}}}_{2}},\ldots,{{{\tilde{u}}}_{K}})}$. The sequence, $\mathbf{u}$, is encoded by a turbo encoder that contains two identical recursive systematic convolutional (RSC) encoders denoted as E1 and E2 as in Fig.~\ref{Fig: turbo_encoder_decoder}. The generator matrix of E1 is ${[1,~{{{g}_{1}} ( D)}/{{{g}_{0}} (D)}]}$, where ${{{g}_{0}}(D)=1+{{D}^{2}}+{{D}^{3}}}$ and ${{{g}_{1}}(D)=1+D+{{D}^{3}}}$~\cite{IEEE3GPP}.
A block of $K$ information bits, ${{u}_{k}}$ for $k=0,~1,~\ldots,~K-1$, is directly passed to the output of the encoder, as the systematic bit sequence ${{\mathbf{x}}^{\rm s}}=(x_{1}^{\rm s},x_{2}^{\rm s},\ldots,x_{K}^{\rm s})=({{u}_{1}},{{u}_{2}},\ldots,{{u}_{K}})$.
E1 generates the parity sequence ${{{\mathbf{x}}^{1\rm p}}=(x_{1}^{1\rm p},x_{2}^{1\rm p},\ldots,x_{K}^{1\rm p})}$ from ${\mathbf{u}}$, and E2 generates the parity sequence ${{{\mathbf{x}}^{2\rm p}}=(x_{1}^{2\rm p},x_{2}^{2\rm p},\ldots,x_{K}^{2\rm p})}$ from $\tilde{\mathbf{u}}$.
Denote ${{S}_{\rm R}}= \{ 0,~1,~\ldots,~7\}$ as the set of all ${{2}^{3}}$ encoder states. ${s\in {S_{\rm R}}}$ is the state of the encoder at time $k$ and ${s'\in {S_{\rm R}}}$ is the state of the encoder at time $k-1$.
The codeword ${{{\mathbf{x}}^{\rm en}}=(x_{1}^{\rm en},x_{2}^{\rm en},\ldots ,x_{K}^{\rm en})}$ consisting of $N=3K$ bits is then modulated and transmitted over channel, where ${x_{i}^{\rm en}=\{x_{i}^{\rm s},x_{i}^{1\rm p},x_{i}^{2\rm p}\}}$.

At the receiver, a soft-output detector computes reliability information in the form of log-likelihood ratios (LLRs) for the transmitted bits, ${{{\mathbf{y}}^{\rm s}}=(y_{1}^{\rm s},y_{2}^{\rm s},\ldots y_{K}^{\rm s})}$, which is the reliability information in the form of LLRs for ${{{\mathbf{x}}^{\rm s}}}$ and indicates the probability of the corresponding bits being 1 or 0. ${{{\mathbf{y}}^{1\rm p}}=(y_{1}^{1\rm p},y_{2}^{1\rm p},\ldots,y_{K}^{1\rm p})}$ and ${{{\mathbf{y}}^{2\rm p}}= (y_{1}^{2\rm p},y_{2}^{2\rm p},\ldots,y_{K}^{2\rm p})}$ are similarly defined for ${{\mathbf{x}}^{1\rm p}}$ and ${{\mathbf{x}}^{2\rm p}}$, respectively. The  LLR sequence ${{\mathbf{y}}^{\rm de}=(y_{1}^{\rm de},y_{2}^{\rm de},\ldots,y_{K}^{\rm de})}$, where ${y_{i}^{\rm de}=\{y_{i}^{\rm s},y_{i}^{\rm 1p},y_{i}^{\rm 2p}\}}$. Denote ${\mathbf{y}_{a}^{b}=({{y}_{a}},{{y}_{a+1}},\ldots,{{y}_{b}})}$, where ${{{y}_{i}}= \{ y_{i}^{\rm s},y_{i}^{1\rm p}\}}$, and then $\mathbf{y}=\mathbf{y}_{1}^{K}=( {{y}_{1}},{{y}_{2}},\ldots,{{y}_{K}})$.

\subsection{Channel and Turbo Decoder}
\label{subsction:the max-log-map algorithm}
The traditional turbo decoder in~\cite{IEEEturbocode:Berrou} contains two soft-input soft-output (SISO) decoders (denoted by D1 and D2) that have the same structure. Therefore, we only introduce D1 in detail as follows. The MAP algorithm is used to compute \emph{a posteriori} LLRs for information bits and can be expressed as:
\begin{equation}
\label{Eq:posteriori llr}
\begin{aligned}
L\left( \left. {{u}_{k}} \right|\mathbf{y} \right)=\log{ \left( \frac{P\left( \left. {{u}_{k}}=1 \right|\mathbf{y} \right)}{P\left( \left. {{u}_{k}}=0 \right|\mathbf{y} \right)} \right)}=\log{\left(\frac{\sum\limits_{\left( {s}',s \right)\in {{S}^{1}}}{P\left( {s}',s,\mathbf{y} \right)}}{\sum\limits_{\left( {s}',s \right)\in {{S}^{0}}}{P\left( {s}',s,\mathbf{y} \right)}}\right)},  
\end{aligned}
\end{equation}
where ${{{S}^{1}}=\{({s}',s):{{u}_{k}}=1\}}$ is the set of ordered pairs ${({s}',s)}$ corresponding to all state transitions ${{s}'\to s}$ caused by data input ${{{u}_{k}}=1}$ and ${{{S}^{0}}=\{({s}',s):{{u}_{k}}=0\}}$ is the set of ordered pairs ${({s}',s)}$ corresponding to all state transitions ${{s}'\to s}$ caused by data input ${{{u}_{k}}=0}$. Let ${S={{S}^{1}}\cup{{S}^{0}}}$ denote the set of ordered pairs ${({s}',s)}$ corresponding to all state transitions ${{s}'\to s}$ whether data input ${{u}_{k}}$ is 1 or 0.
A total of 16 possible state transitions in set $S$ and the corresponding parity bit ${x_{k}^{1\rm p}}$ for input bit ${{u}_{k}}$ in E1 are presented in the Table~\ref{tab:state transitions}. A critical step in the turbo decoder is to obtain \emph{a posteriori} LLRs for information bits. In the following, we will present the max-log-MAP algorithm~\cite{IEEEReduced:Erfanian}.
\begin{table}[t]
	\caption{State transitions of RSC encoder}
	\label{tab:state transitions}
	\centering
	\begin{tabular}{c|ccccccccc}
		\toprule
		& ${s}'$& 0 & 1 & 2 & 3 & 4 & 5 & 6 & 7  \\
		\hline
		\multirow{2}{*}{\rotatebox{0}{${{u}_{k}=0}$}} & ${s}$ & 0 & 4 & 5 & 1 & 2 & 6 & 7 & 3\\
		& $x_{k}^{1\rm p}$ & 0 & 0 & 1 & 1 & 1 & 1 & 0 & 0\\
		\hline
		\multirow{2}{*}{\rotatebox{0}{${{u}_{k}=1}$}} & ${s}$ & 4 & 0 & 1 & 5 & 6 & 2 & 3 & 7\\
		& $x_{k}^{1\rm p}$ & 1 & 1 & 0 & 0 & 0 & 0 & 1 & 1\\
		\bottomrule
	\end{tabular}
\end{table}

Assume that the additive white Gaussian noise (AWGN) channel is memoryless . Then, the sequence received after time $k$ is only related to the state of the encoder at time $k$ regardless of the previous states. On the basis of the Bayes formula, we obtain
\begin{equation}
\label{Eq:p(s',s,y)}
\begin{aligned}
P\left( {s}',s,\mathbf{y} \right)&=P\left( {s}',s,\mathbf{y}_{1}^{k-1},{{y}_{k}},\mathbf{y}_{k+1}^{K} \right)\\
&=P\left( \left. \mathbf{y}_{k+1}^{K} \right|s \right)P\left( \left. {{y}_{k}},s \right|{s}' \right)P\left( {s}',\mathbf{y}_{1}^{k-1} \right)\\
&={{\beta }_{k}}\left( s \right){{\gamma }_{k}}\left( {s}',s \right){{\alpha }_{k-1}}\left( {{s}'} \right),  
\end{aligned}
\end{equation}
where ${{{\alpha }_{k-1}}( {{s}'})=P({s}',\mathbf{y}_{1}^{k-1})}$ and ${{{\beta }_{k}}( s)=P(\mathbf{y}_{k+1}^{K}|s)}$ can be computed through the forward and backward recursions~\cite{IEEEoptimaldecoding:Bahl}
\begin{equation}
\label{Eq:alpha}
\begin{aligned}
{{\alpha }_{k}}\left( s \right)=\sum\limits_{s'\in {{S}_{\rm R}}}{{{\alpha }_{k-1}}\left(s' \right){{\gamma }_{k}}\left( s',s \right)},
\end{aligned}
\end{equation}
and
\begin{equation}
\label{Eq:beta}
\begin{aligned}
{{\beta }_{k-1}}\left(s' \right)=\sum\limits_{s\in {{S}_{\rm R}}}{{{\beta }_{k}}\left( s \right){{\gamma }_{k}}\left( s',s \right)}
\end{aligned}
\end{equation}
with initial conditions ${{{\alpha }_{0}}(0)=1}$, ${{{\alpha }_{0}}(n)=0}$ for $n\ne 0$, and ${{{\beta }_{K}} (0)=1}$, ${{{\beta}_{K}}(n)=0}$ for $n\ne 0$. (The encoder is expected to end in state 0 after $K$ input bits, implying that the last three input bits, called termination bits, are so selected.) Moreover, ${{\gamma}_{k}}(s',s)=P({{y}_{k}},s|s')$ is computed as follows~\cite{IEEEiterative:Bauch}:
\begin{equation}
\label{Eq:gamma2}
\begin{aligned}
{{\gamma }_{k}}\left( {s}',s \right)=\exp \left\{ \frac{1}{2}\left( x_{k}^{\rm s}y_{k}^{\rm s}+x_{k}^{1\rm p}y_{k}^{1\rm p} \right)+\frac{1}{2}{{u}_{k}}L\left( {{u}_{k}} \right) \right\},
\end{aligned}
\end{equation}
where $L\left( {{u}_{k}} \right)$ is the \emph{a priori} probability LLR for bit ${{u}_{k}}$. Given that $L(\left. {{u}_{k}} \right|\mathbf{y})$ is the sum of the systematic bit LLR $y_{k}^{\rm s}$, the \emph{a priori} probability LLR $L({{u}_{k}})$, and the extrinsic LLR ${{L}_{\rm e}}\left( {{u}_{k}} \right)$, we obtain
\begin{equation}
\label{Eq:le}
\begin{aligned}
{{L}_{\rm e}}\left( {{u}_{k}} \right)=L(\left. {{u}_{k}} \right|\mathbf{y})-y_{k}^{\rm s}-L({{u}_{k}}),
\end{aligned}
\end{equation}
which can be used as the \emph{a priori} probability LLR input of the subsequent SISO D2 after it is interleaved.

Let ${{\bar{\alpha }}_{k}}\left( s \right)$, ${{\bar{\beta }}_{k}}\left( s \right)$, and ${{\bar{\gamma }}_{k}}\left( s',s \right)$ represent the logarithmic values of ${{\alpha }_{k}}\left( s \right)$, ${{\beta }_{k}}\left( s \right)$, and ${{\gamma }_{k}}\left( s',s \right)$, respectively. The log-MAP algorithm~\cite{IEEEOptimal:Robertson} evaluates ${{{\alpha }_{k-1}}(s')}$ and ${{{\beta}_{k}}(s)}$ in logarithmic terms using the Jacobian logarithmic function:
\begin{equation}
\label{Eq:log_alpha}
\begin{aligned}
{{\bar{\alpha }}_{k}}\left( s \right)=\underset{{s}'\in {{S}_{\rm R}}}{\mathop{\max }}\,{{}^{*}}\left( {{{\bar{\alpha }}}_{k-1}}\left( {{s}'} \right)+{{{\bar{\gamma }}}_{k}}\left( {s}',s \right) \right)
\end{aligned}
\end{equation}
and
\begin{equation}
\label{Eq:log_beta}
\begin{aligned}
{{\bar{\beta }}_{k-1}}\left( {{s}'} \right)=\underset{s\in {{S}_{\rm R}}}{\mathop{{{\max }^{*}}}}\,\left( {{{\bar{\beta }}}_{k}}\left( s \right)+{{{\bar{\gamma }}}_{k}}\left({s}',s\right)\right),
\end{aligned}
\end{equation}
where ${{\max}^{*}}(x,y)=\max(x,y)+\log( 1+{{e}^{-| x-y|}})$,
and initial conditions ${{\bar{{\alpha }}_{0}}(0)=0}$, ${{\bar{{\alpha }}_{0}}(n)=-\infty}$ for $n\ne 0$, and ${{\bar{{\beta }}_{K}} (0)=0}$, ${{\bar{{\beta}}_{K}}(n)=-\infty}$ for $n\ne 0$ (A large negative number, e.g.,~--128, will be used for programming instead of minus infinity). The \emph{a posteriori} LLRs for information bits are computed by
\begin{equation}
\label{Eq:posteriori llr2}
\begin{aligned}
L\left( \left. {{u}_{k}} \right|\mathbf{y} \right)&=\underset{\left( {s}',s \right)\in {{S}^{1}}}{\mathop{{{\max }^{*}}}}\,\left( {{{\bar{\alpha }}}_{k-1}}\left( {{s}'} \right)+{{{\bar{\gamma }}}_{k}}\left( {s}',s \right)+{{{\bar{\beta }}}_{k}}\left( s \right) \right)\\
&-\underset{\left( {s}',s \right)\in {{S}^{0}}}{\mathop{{{\max }^{*}}}}\,\left( {{{\bar{\alpha }}}_{k-1}}\left( {{s}'} \right)+{{{\bar{\gamma }}}_{k}}\left( {s}',s \right)+{{{\bar{\beta }}}_{k}}\left( s \right) \right).
\end{aligned}
\end{equation}

The max-log-MAP algorithm~\cite{IEEEReduced:Erfanian} omits the logarithmic term in the Jacobian logarithmic function. Hence, equations (\ref{Eq:log_alpha})--(\ref{Eq:posteriori llr2}) can be approximately written as:
\begin{equation}
\label{Eq:max log alpha}
\begin{aligned}
{{\bar{\alpha }}_{k}}\left( s \right)=\underset{{s}'\in {{S}_{\rm R}}}{\mathop{\max }}\,\left( {{{\bar{\alpha }}}_{k-1}}\left( {{s}'} \right)+{{{\bar{\gamma }}}_{k}}\left( {s}',s \right) \right),
\end{aligned}
\end{equation}
\begin{equation}
\label{Eq:max log beta}
\begin{aligned}
{{\bar{\beta }}_{k-1}}\left( {{s}'} \right)=\underset{s\in {{S}_{\rm R}}}{\mathop{\max }}\,\left( {{{\bar{\beta }}}_{k}}\left( s \right)+{{{\bar{\gamma }}}_{k}}\left( {s}',s \right) \right),
\end{aligned}
\end{equation}
and
\begin{equation}
\label{Eq:max_posteriori llr2}
\begin{aligned}
L\left( \left. {{u}_{k}} \right|\mathbf{y} \right)&=\underset{\left( {s}',s \right)\in {{S}^{1}}}{\mathop{\max }}\,\left( {{{\bar{\alpha }}}_{k-1}}\left( {{s}'} \right)+{{{\bar{\gamma }}}_{k}}\left( {s}',s \right)+{{{\bar{\beta }}}_{k}}\left( s \right) \right)\\
&-\underset{\left( {s}',s \right)\in {{S}^{0}}}{\mathop{\max }}\,\left( {{{\bar{\alpha }}}_{k-1}}\left( {{s}'} \right)+{{{\bar{\gamma }}}_{k}}\left( {s}',s \right)+{{{\bar{\beta }}}_{k}}\left( s \right) \right).
\end{aligned}
\end{equation}

\section{Basic DNN Decoder: TurboNet}
\label{section: TurboNet Architecture}
The max-log-MAP algorithm~\cite{IEEEReduced:Erfanian} introduced in the previous section calculates \emph{a posteriori} LLRs iteratively, which fits well for model-driven DL framework. This section elaborates the integration of DL into the traditional MAP algorithm, that is, how to obtain an alternative graphical representation to replace the traditional max-log-MAP algorithm and introduce trainable parameters. The TurboNet inherits the superiority of the max-log-MAP algorithm and DL techniques, and thus presents excellent error-correction ability with excessively low training cost.

\begin{figure}[tb]
	\centering
	\includegraphics[width=5in]{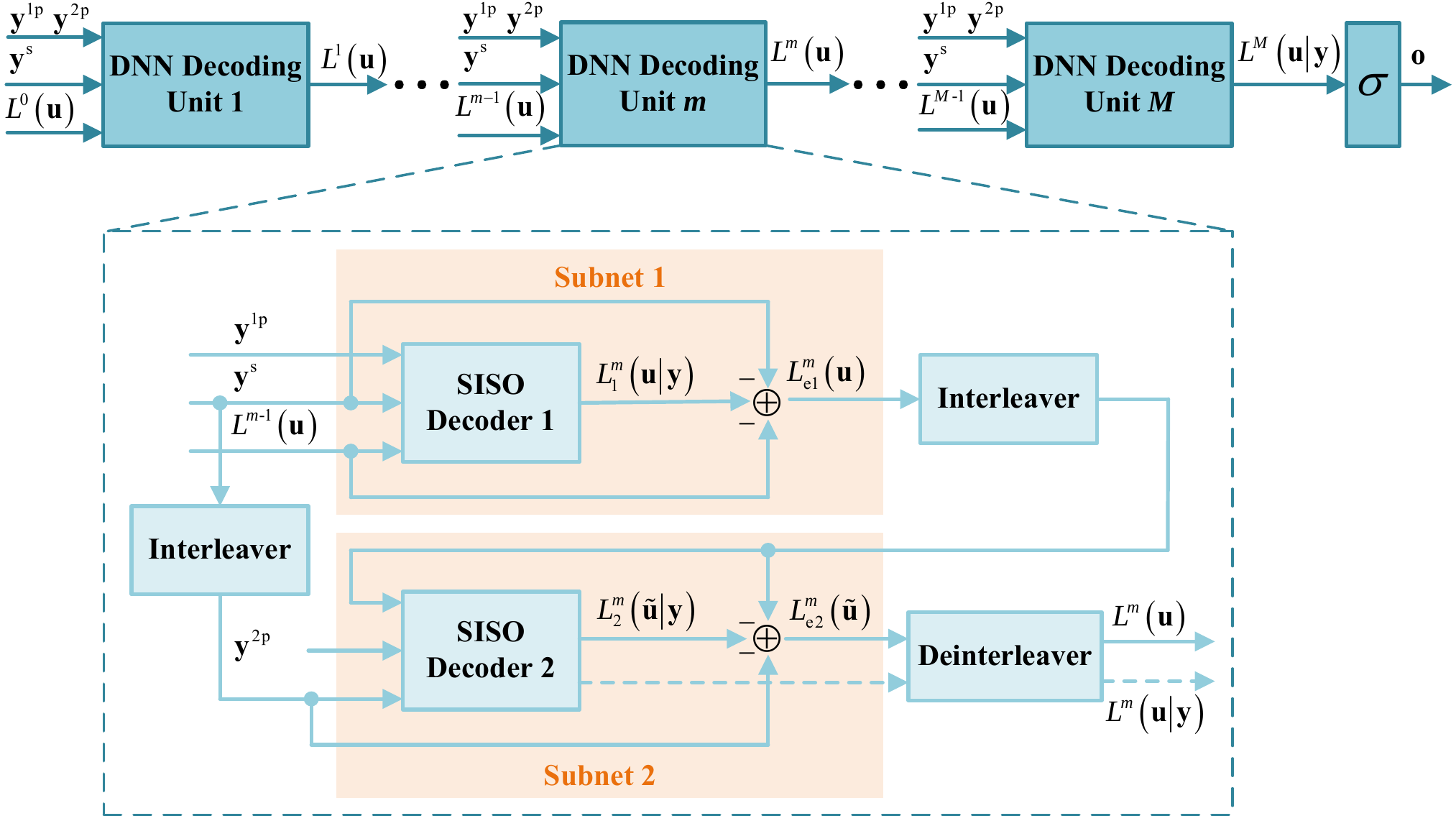}
	\caption{The TurboNet architecture. Each DNN decoding unit stands for one iteration. The output of the DNN decoding unit $M$ is ${{{L}^{M}}({\mathbf{u}}|\mathbf{y})}$ rather than ${{{L}^{M}}({\mathbf{u}})}$, and \emph{a priori} probability LLRs ${{L}^{0}}( {{u}_{k}})$,  the input of DNN decoding unit 1, are initialized to ${\mathbf{0}}$, i.e., ${{{L}^{0}}({{u}_{k}})=0}$ for $k=1,~2,~\ldots,~K$. The dotted arrow from the deinterleaver denotes the output of the DNN decoding unit $m$ for ${m=M}$, otherwise, the solid arrow from the deinterleaver represents the output.}
	\label{Fig:dnn_turbo_decoder}
\end{figure}

The traditional iterative structure is first unfolded and each iteration is represented by a DNN decoding unit to obtain an ``unrolled'' structure as in~Fig.~\ref{Fig:dnn_turbo_decoder}, which is equivalent to $M$ iterations. For each iteration, conventional SISO decoders using the max-log-MAP algorithm and the part of calculating the extrinsic LLRs are replaced by two subnets based on the neural max-log-MAP algorithm. That is, we \emph{integrate DNN} into \emph{the max-log-MAP algorithm} but preserve the original architecture, or alternatively, the DNN decoding unit $m$ contains two identical interleavers, two subnets with an identical structure but no shared parameters, and one deinterleaver. ${{{L}^{m}} ({\mathbf{u}})}$ denotes the \emph{a priori} probability LLRs calculated by the max-log-MAP algorithm with $m$ iterations, and ${{{L}^{M}}({\mathbf{u}}|\mathbf{y})}$ is \emph{a posteriori} LLRs calculated by the max-log-MAP algorithm with $M$ iterations, where ${
	 m=0,~1,~\dots,~M-1}$. For D1, ${L_{1}^{m}({\mathbf{u}}|\mathbf{y})}$ and ${L_{{\rm e}1}^{m}( {\mathbf{u}})}$ represent the \emph{a posteriori} probability LLRs and the extrinsic LLRs calculated by the max-log-MAP algorithm with $m$ iterations, respectively. ${\mathbf{o}=({{o}_{1}},~{{o}_{2}},~\ldots,~{{o}_{K}})}$, where ${{o}_{k}}$ indicates the estimated probability of bit ${{u}_{k}}$ being a binary 1.

Subnet 1 (SN1), which is based on a neural max-log-MAP algorithm, contains one input layer, $K+1$ hidden layers, and one output layer. Subnet 2 (SN2) has the same structure as SN1. The details of the SN1 architecture in the DNN decoding unit $m$ are elaborated as follows:
\subsection{Input Layer}
The input layer of the proposed subnet consists of $3K$ neurons and the input of the neurons constitutes set ${{{N}^{\rm In}}=\{{{I}_{k}}:k=1,~2,~\ldots,~K\}}$, where the triplet ${{{I}_{k}}=\{y_{k}^{\rm s},~y_{k}^{1\rm p},~{{L}^{m-1}}({{u}_{k}})\}}$.
\subsection{Hidden Layer 1}
\begin{figure}[tb]
	\centering
	\includegraphics[width=2.5in]{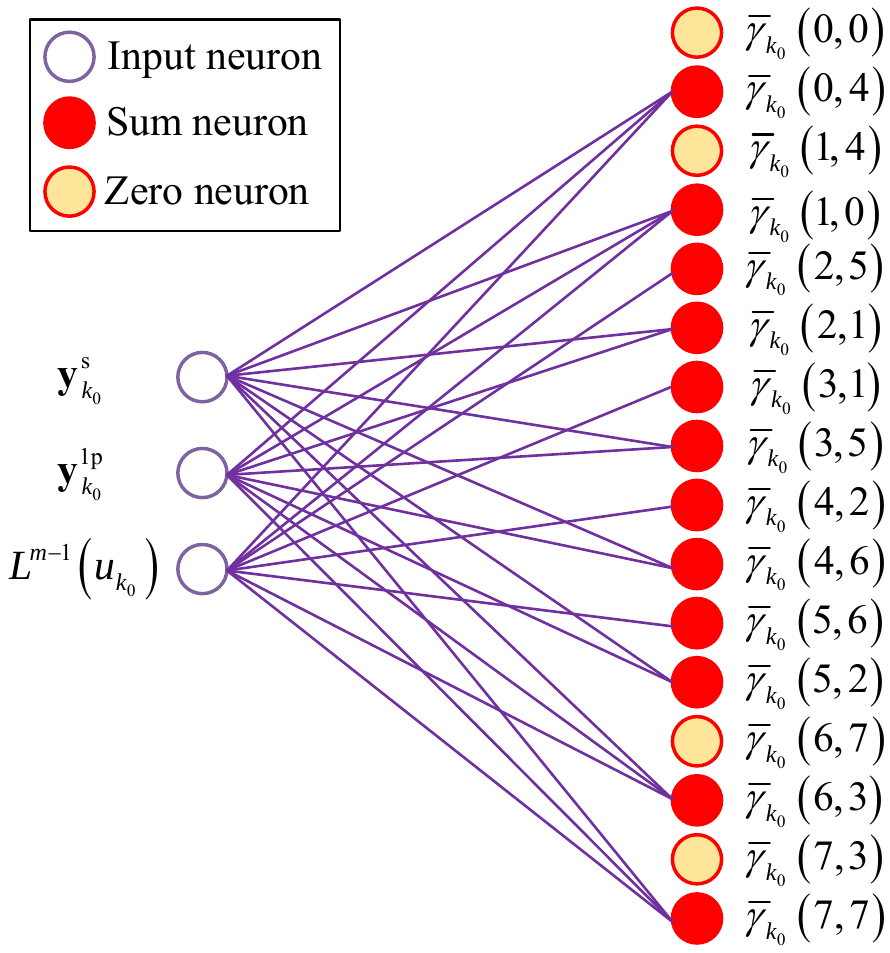}
	\caption{Hidden layer 1 architecture. The input neuron represents the input value of the network. The sum neuron implements a weighted summation of the input data without an activation function. The zero neuron represents a constant output value of 0.}
	\label{Fig:hidden_layer_1_nonfc}
\end{figure}

The first hidden layer contains $16K$ neurons and the output of the neurons in this layer constitutes set ${{{N}^{1}}= \{{{{\bar{\gamma}}}_{k}}({s}',s):({s}',s)\in S,k=1,~2,~\ldots ,~K\}}$. The neuron corresponding to ${{{\bar{\gamma}}_{{{k}_{0}}}}({{s}'_{0}},{{s}_{0}})\in {{N}^{1}}}$ in the first hidden layer is connected to the neurons that correspond to ${y_{{{k}_{0}}}^{\rm s}}$, ${y_{{{k}_{0}}}^{1\rm p}}$, and ${{{L}^{m-1}}({{u}_{k}})}$ in the input layer, where ${({{s}'_{0}},{{s}_{0}})\in S}$ and ${{{k}_{0}}\in \{ 1,~2,~\ldots ,~K \}}$. Fig.~\ref{Fig:hidden_layer_1_nonfc} shows the connections between the neurons corresponding to the elements in set ${\{{{\gamma}_{{{k}_{0}}}}({s}',s):({s}',s)\in S\}}$ in hidden layer~1 and the neurons corresponding to ${y_{{{k}_{0}}}^{\rm s}}$, ${y_{{{k}_{0}}}^{1\rm p}}$, and ${{{L}^{m-1}}({{u}_{k}})}$ in the input layer.
Notably, not all neurons in the first hidden layer need to be connected to the input neurons. For example, Table~\ref{tab:state transitions} shows that if ${{{s}'_{0}}=2}$ and ${{{u}_{{{k}_{0}}}}=x_{{{k}_{0}}}^{\rm s}=0}$, then ${x_{{{k}_{0}}}^{1\rm p}=1}$, and we obtain
\begin{equation}
\label{Eq:weight gamma example 1}
\begin{aligned}
{{\bar{\gamma }}_{{{k}_{0}}}}\left( 2,5 \right)=\frac{1}{2}w_{\bar{\gamma },{{k}_{0}},\left( 2,5 \right)}^{3}y_{{{k}_{0}}}^{1\rm p},
\end{aligned}
\end{equation}
which means that the neuron corresponding to ${{{\bar{\gamma }}_{{{k}_{0}}}}(2,5)}$ in hidden layer 1 is only connected to the neuron corresponding to ${y_{{{k}_{0}}}^{1\rm p}}$ in the input layer. Similarly, we can obtain
\begin{equation}
\label{Eq:weight gamma example 2}
\begin{aligned}
{{\bar{\gamma }}_{{{k}_{0}}}}\left( 2,1 \right)=\frac{1}{2}w_{\bar{\gamma },\left( 2,1 \right),{{k}_{0}}}^{1}{{L}^{m-1}}\left( {{u}_{{{k}_{0}}}} \right)+\frac{1}{2}w_{\bar{\gamma },\left( 2,1 \right),{{k}_{0}}}^{2}y_{{{k}_{0}}}^{\rm s},
\end{aligned}
\end{equation}
and
\begin{equation}
\label{Eq:weight gamma example 3}
\begin{aligned}
{{\bar{\gamma }}_{{{k}_{0}}}}\left( 0,0 \right)=0.
\end{aligned}
\end{equation}
Identity (\ref{Eq:weight gamma example 3}) shows that the neuron corresponding ${{{\bar{\gamma }}_{{{k}_{0}}}}(0,0)}$ is not connected to any neuron in the input layer and can be considered a constant output value of 0. Accordingly, we can obtain a partially connected structure as shown in Fig.~\ref{Fig:hidden_layer_1_nonfc}.
We assign weights to the edges between the neurons in the input layer and hidden layer 1 shown in Fig.~\ref{Fig:hidden_layer_1_nonfc}. The neuron corresponding to ${{\bar{\gamma}}_{{{k}_{0}}}}({{s}'_{0}},{{s}_{0}})$ calculates the output as follows:
\begin{equation}
\label{Eq:weigh gamma}
\begin{aligned}
{{\bar{\gamma }}_{{{k}_{0}}}}\left( {{s}'_{0}},{{s}_{0}} \right)=\frac{1}{2}w_{\bar{\gamma },\left( {{s}'_{0}},{{s}_{0}} \right),{{k}_{0}}}^{1}\left\{ {{u}_{{{k}_{0}}}}{{L}^{m-1}}\left( {{u}_{{{k}_{0}}}} \right) \right\}+\frac{1}{2}w_{\bar{\gamma },\left( {{s}'_{0}},{{s}_{0}} \right),{{k}_{0}}}^{2}\left\{ x_{{{k}_{0}}}^{\rm s}y_{{{k}_{0}}}^{\rm s} \right\}+\frac{1}{2}w_{\bar{\gamma },\left( {{{{s}'}}_{0}},{{s}_{0}} \right),{{k}_{0}}}^{3}\left\{ x_{{{k}_{0}}}^{1\rm p}y_{{{k}_{0}}}^{1\rm p} \right\}.
\end{aligned}
\end{equation}

\subsection{Hidden Layer from 2 to K}
Each layer of the following $K-1$ hidden layers contains ${2\times| {{S}_{\rm R}}|=16}$ neurons. For the $z$th hidden layer, the output of all neurons constitutes the set ${{N}^{z}}=N_{\rm odd}^{z}\cup N_{\rm even}^{z}$, where ${N_{\rm odd}^{z}=\{{{{\bar{\alpha }}}_{k}}(s):k=z-1,s\in{{S}_{\rm R}}\}}$ is the set of neuron outputs for all odd positions in the $z$th hidden layer, ${N_{\rm even}^{z}= \{{{{\bar{\beta}}}_{k-1}}({s}'):k=K-z+2,{s}'\in {{S}_{\rm R}}\}}$ is the set of neuron outputs for all even positions in the $z$th hidden layer, for $z=2,~3,~\ldots,~K$. The hidden layers 2 to $K$ calculate ${{{{\bar{\alpha }}}_{k}}(s)}$ and ${{{{\bar{\beta}}}_{k-1}}({s}')}$ according to~(\ref{Eq:max log alpha}) and~(\ref{Eq:max log beta}), and details are elaborated as follows:

\begin{figure}[t]
	\centering
	\subfigure[Hidden layer ${{z}_{0}}$ architecture (odd position)]{
		\begin{minipage}[t]{0.45\textwidth}
			\centering
			\includegraphics[width=2.1in]{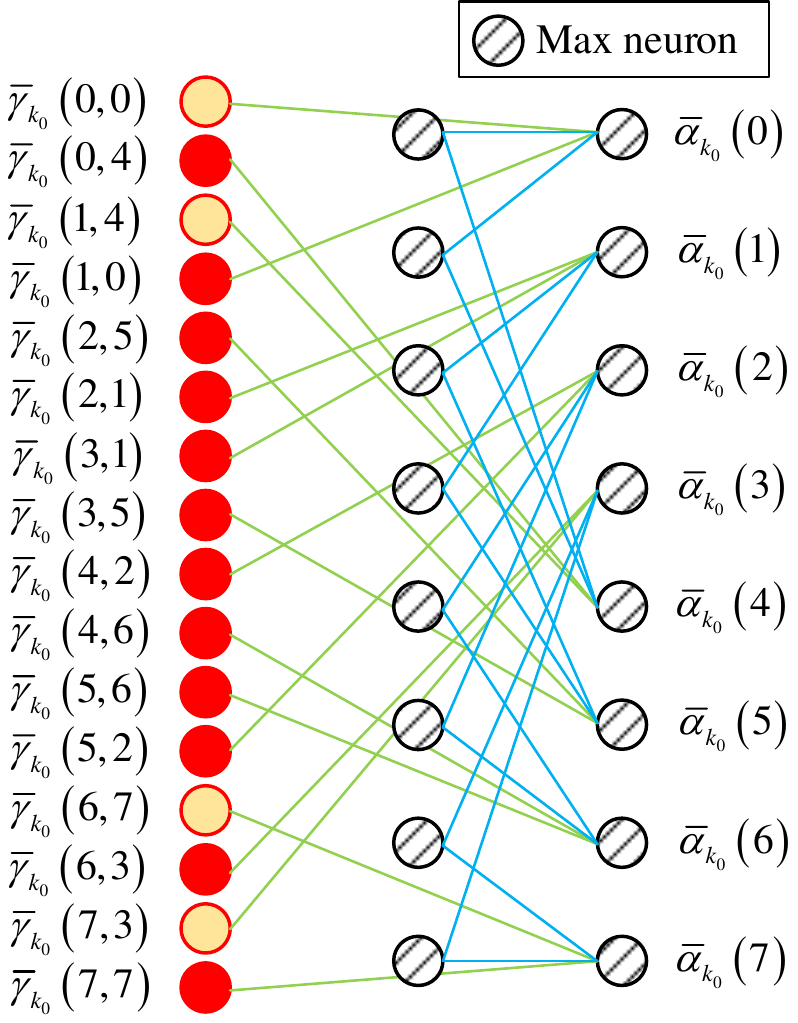}
			\label{Fig:hidden_layer_3_later_odd}
	\end{minipage}}
	\subfigure[Hidden layer ${{z}_{0}}$ architecture (even position)]{
		\begin{minipage}[t]{0.45\textwidth}
			\centering
			\includegraphics[width=2.1in]{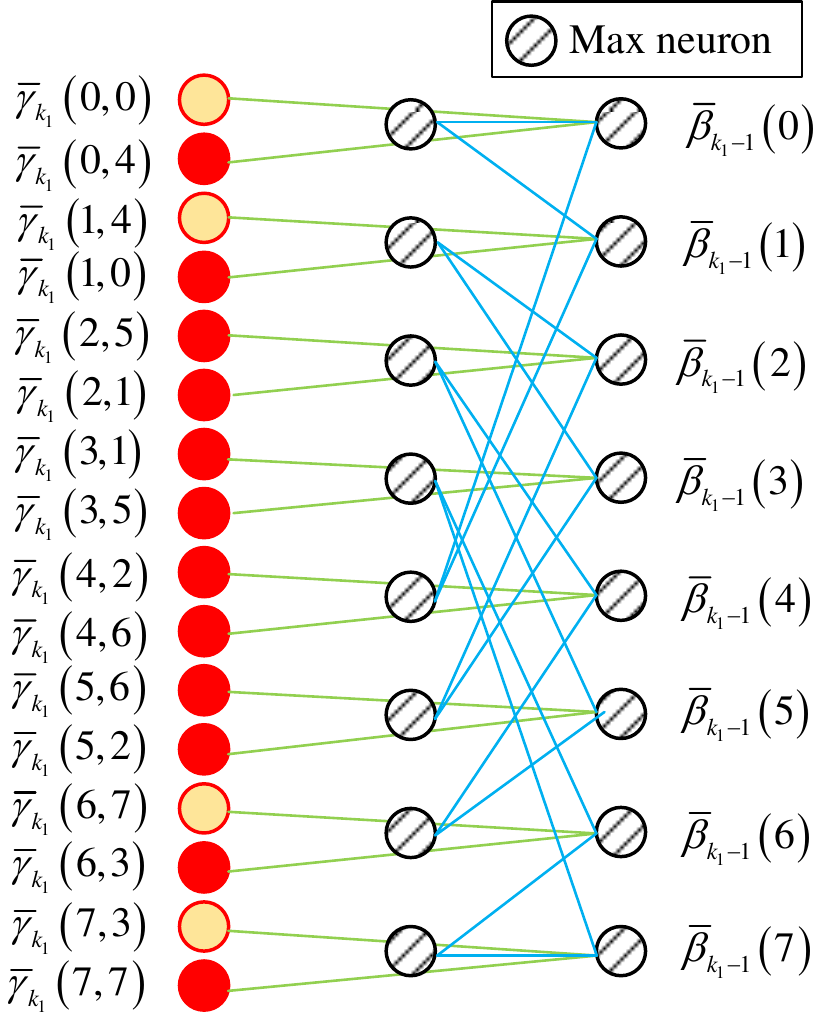}
			\label{Fig:hidden_layer_3_later_even}
	\end{minipage}}	
	\caption{Hidden layer ${{z}_{0}}$ architecture: the (a) odd position and (b) even position. The max neuron calculates the maximum sum of the input groups expressed as (\ref{Eq:weight alpha}) and (\ref{Eq:weight beta}).}
\end{figure}
\subsubsection{For ${{{z}_{0}}\in \{3,~4,~\ldots ,~K\}}$}
The neuron corresponding to ${{{\bar{\alpha}}_{{{k}_{0}}}}({{s}_{0}})\in N_{\rm odd}^{{{z}_{0}}}}$ in the ${{z}_{0}}$th layer is connected to all neurons corresponding to the elements in set ${\{{{{\bar{\alpha }}}_{{{k}_{0}}-1}} ( {{s}'}):( {s}',{{s}_{0}})\in S \}}$ in layer ${{z}_{0}-1}$ and all neurons corresponding to the elements in set ${\{{{{\bar{\gamma}}}_{{{k}_{0}}}} ( {s}',{{s}_{0}}): ( {s}',{{s}_{0}})\in S \}}$ in hidden layer~1, where ${{{k}_{0}}={{z}_{0}}-1}$ and ${{{s}_{0}}\in {{S}_{\rm R}}}$. The neuron corresponding to ${{{\bar{\beta }}_{{{k}_{1}}-1}}({{s}'_{0}})\in N_{\rm even}^{{{z}_{0}}}}$ in the ${{z}_{0}}$th layer is connected to the neurons corresponding to the elements in set ${\{{{{\bar{\beta }}}_{{{k}_{1}}}}(s):({{s}'_{0}},s)\in S\}}$ in layer ${{z}_{0}-1}$ and all neurons corresponding to the elements in set ${\{{{{\bar{\gamma}}}_{{{k}_{1}}}} ({s}'_{0},s):({s}'_{0},s)\in S \}}$ in the first hidden layer, where ${{{k}_{1}}=K-{{z}_{0}}+2}$ and ${{{s}'_{0}}\in {{S}_{\rm R}}}$. Figs.~\ref{Fig:hidden_layer_3_later_odd} and~\ref{Fig:hidden_layer_3_later_even}  show the connections between all neurons in the ${{z}_{0}}$th hidden layer and the neurons in the previous hidden layers (i.e., hidden layer 1 and hidden layer ${{z}_{0}-1}$).

Turbo codes usually have a large block size. For example, the minimum message bit length of turbo codes in the long-term evolution standard is 40, and the maximum is 6144. Therefore, parameterizing (\ref{Eq:max log alpha}) and (\ref{Eq:max log beta}) will cause the neural network in Fig.~\ref{Fig:dnn decoder} to be extremely ``deep'' and may lead to gradient vanishing or gradient exploding. Therefore, we will not introduce any trainable parameters in these layers. The neurons corresponding to ${{{\bar{\alpha }}_{{{k}_{0}}}}({{s}_{0}})}$ and ${{{\bar{\beta }}_{{{k}_{1}}-1}}({{s}'_{0}})}$ calculate the output as follows:
\begin{equation}
\label{Eq:weight alpha}
\begin{aligned}
{{\bar{\alpha }}_{{{k}_{0}}}}\left( {{s}_{0}} \right)=\underset{s'\in {S}_{\rm R}}{\mathop{\max }}\,\left( {{{\bar{\alpha }}}_{{{k}_{0}}-1}}\left( {{s}'} \right)+{{{\bar{\gamma }}}_{{{k}_{0}}}}\left( {s}',{{s}_{0}} \right) \right)
\end{aligned}
\end{equation}
and
\begin{equation}
\label{Eq:weight beta}
\begin{aligned}
{{\bar{\beta }}_{{{k}_{1}}-1}}\left( {{s}'_{0}} \right)=\underset{s\in {S}_{\rm R}}{\mathop{\max }}\,\left( {{{\bar{\beta }}}_{{{k}_{1}}}}\left( s \right)+{{{\bar{\gamma }}}_{{{k}_{1}}}}\left( {{s}'_{0}},s \right) \right).
\end{aligned}
\end{equation}
\subsubsection{$z=2$}
Given that the initial conditions ${{\bar{{\alpha }}_{0}}(0)=0}$, ${{\bar{{\alpha }}_{0}}(n)=-128}$ for $n\ne 0$, and ${{\bar{{\beta }}_{K}} (0)=0}$, ${{\bar{{\beta}}_{K}}(n)=-128}$ for $n\ne 0$, the neurons in hidden layer 2 are connected to hidden layer 1 and some constant neurons. We employ a small bold line with a black circle to represent a constant input value of --128 in Fig.~\ref{Fig:dnn decoder}.

\subsection{Hidden Layer $K+1$}
Hidden layer $K+1$, which is the last hidden layer, consists of $K$ neurons, and the output of all neurons constitutes set ${{{N}^{K+1}}=\{L_{1}^{m}({{u}_{k}}|\mathbf{y}):k=1,~2,~\ldots ,~K\}}$. The neuron corresponding to ${L_{1}^{m}({{u}_{{{k}_{0}}}}|\mathbf{y})\in {{N}^{K+1}}}$ in the last hidden layer is connected to all neurons corresponding to the elements in set ${ \{ {{{\bar{\alpha }}}_{{{k}_{0}}-1}} ( {{s}'}):{s}'\in {{S}_{\rm R}} \}}$, $ {\{ {{{\bar{\gamma }}}_{{{k}_{0}}}} ( {s}',s) : ( {s}',s)\in S \}}$, and ${ \{ {{{\bar{\beta }}}_{{{k}_{0}}}} ( s) :s\in {{S}_{\rm R}} \}}$, where ${{{k}_{0}}\in \{1,~2,~\ldots ,~K\}}$. Fig.~\ref{Fig: hidden_layer_K+1} illustrates the connections between the neuron corresponding to ${ L_{1}^{m}({{u}_{{{k}_{0}}}}|\mathbf{y})} $ in the last hidden layer and the neurons in the previous hidden layers (i.e., hidden layer 1, hidden layer ${{k}_{0}}$, and hidden layer ${K-{{k}_{0}+1}}$).

We assign weights to the edges in Fig.~\ref{Fig: hidden_layer_K+1}. The neuron corresponding to ${ L_{1}^{m}({{u}_{{{k}_{0}}}}|\mathbf{y})} $ calculates the output as follows:
\begin{figure}[tb]
	\centering
	\includegraphics[width=3.2in]{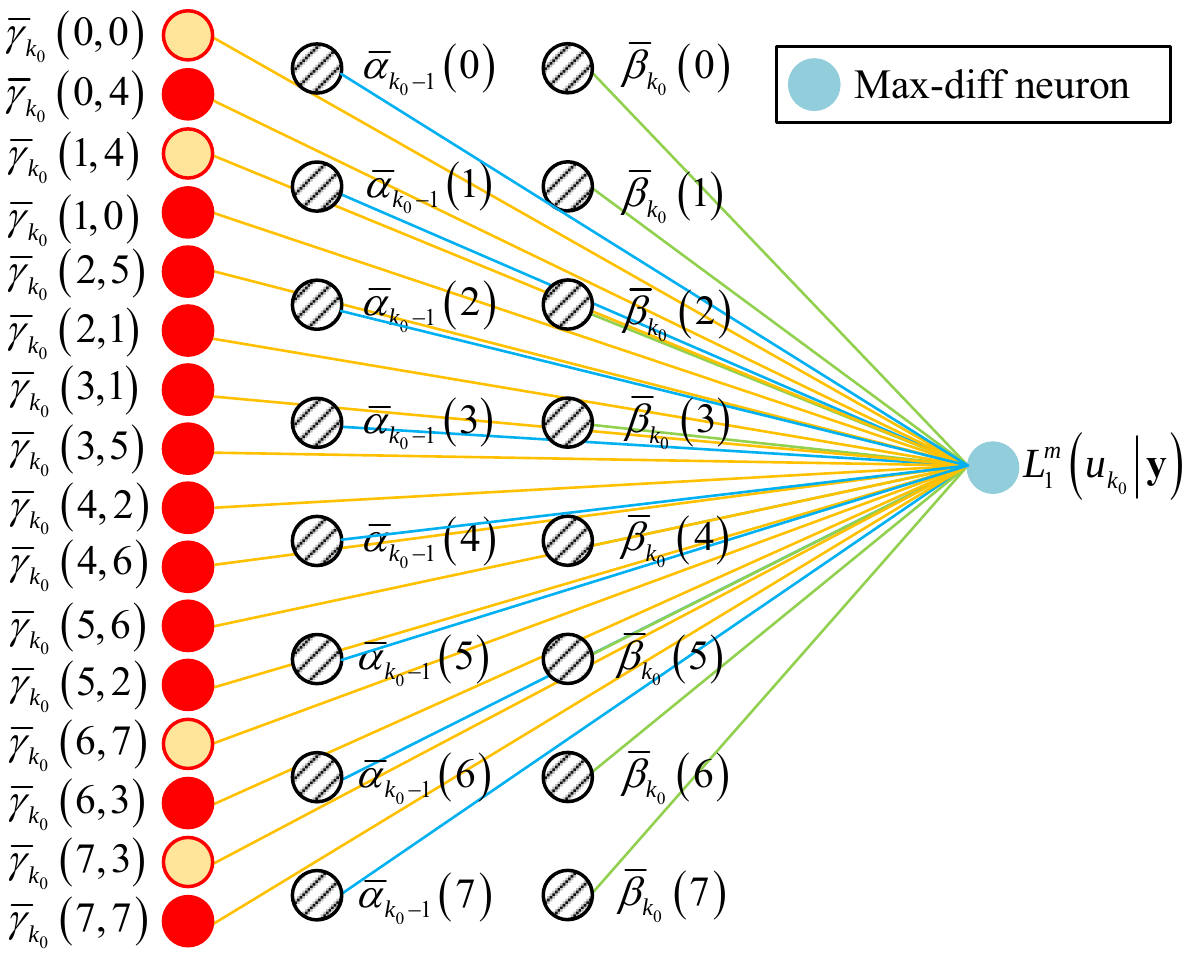}
	\caption{Hidden layer ${K+1}$ architecture. The max-diff neuron realizes that the input data are divided into two categories, and the maximum values of the sum of the groups in both categories are obtained to calculate the difference as shown by (\ref{Eq:weigh posteriore llr}).}
	\label{Fig: hidden_layer_K+1}
\end{figure}

\begin{equation}
    \label{Eq:weigh posteriore llr}
    \begin{aligned}
    L_{1}^{m}\left( \left. {{u}_{{{k}_{0}}}} \right|\mathbf{y} \right)&=\underset{\left( s',s \right)\in {{S}^{1}}}{\mathop{\max }}\,\left( w_{{s}',{{k}_{0}}}^{1}{{{\bar{\alpha }}}_{{{k}_{0}}-1}}\left( s' \right)+w_{\left( {s}',s \right),{{k}_{0}}}^{2}{{{\bar{\gamma }}}_{{{k}_{0}}}}\left( s',s \right)+w_{s,{{k}_{0}}}^{3}{{{\bar{\beta }}}_{{{k}_{0}}}}\left( s \right) \right) \\ 
    & -\underset{\left( s',s \right)\in {{S}^{0}}}{\mathop{\max }}\,\left( w_{{s}',{{k}_{0}}}^{4}{{{\bar{\alpha }}}_{{{k}_{0}}-1}}\left( s' \right)+w_{\left( {s}',s \right),{{k}_{0}}}^{5}{{{\bar{\gamma }}}_{{{k}_{0}}}}\left( s',s \right)+w_{s,{{k}_{0}}}^{6}{{{\bar{\beta }}}_{{{k}_{0}}}}\left( s \right) \right).
    \end{aligned}
\end{equation}

\subsection{Output Layer}
The output layer contains $K$ neurons, and the output of all neurons in this layer constitutes set ${{{N}^{{\rm Out}}}=\{L_{{\rm e}1}^{m}({{u}_{k}}):k=1,~2,~\ldots ,~K\}}$. The neuron corresponding to ${L_{{\rm e}1}^{m}({{u}_{{{k}_{0}}}})\in {{N}^{{\rm Out}}}}$ is connected to the neuron corresponding to ${ L_{1}^{m}({{u}_{{{k}_{0}}}}|\mathbf{y})} $ in hidden layer $K+1$ and the neurons corresponding to ${y_{{{k}_{0}}}^{\rm s}}$, ${{{L}^{m-1}}({{u}_{{{k}_{0}}}})}$ in the input layer. 

We assign weights to the edges connected to the neuron corresponding to ${L_{{\rm e}1}^{m}({{u}_{{{k}_{0}}}})}$ in the output layer and this neuron calculates the output as follows:
\begin{equation}
\label{Eq:weigh extrinsic llr}
\begin{aligned}
L_{{\rm e}1}^{m}\left( {{u}_{{{k}_{0}}}} \right)=w_{{\rm e},{{k}_{0}}}^{1}L_{1}^{m}(\left. {{u}_{{{k}_{0}}}} \right|\mathbf{y})-w_{{\rm e},{{k}_{0}}}^{2}y_{{{k}_{0}}}^{\rm s}-w_{{\rm e},{{k}_{0}}}^{3}{{L}^{m-1}}({{u}_{{{k}_{0}}}}).  
\end{aligned}
\end{equation}

\begin{figure}[t]
	\centering
	\includegraphics[width=3in]{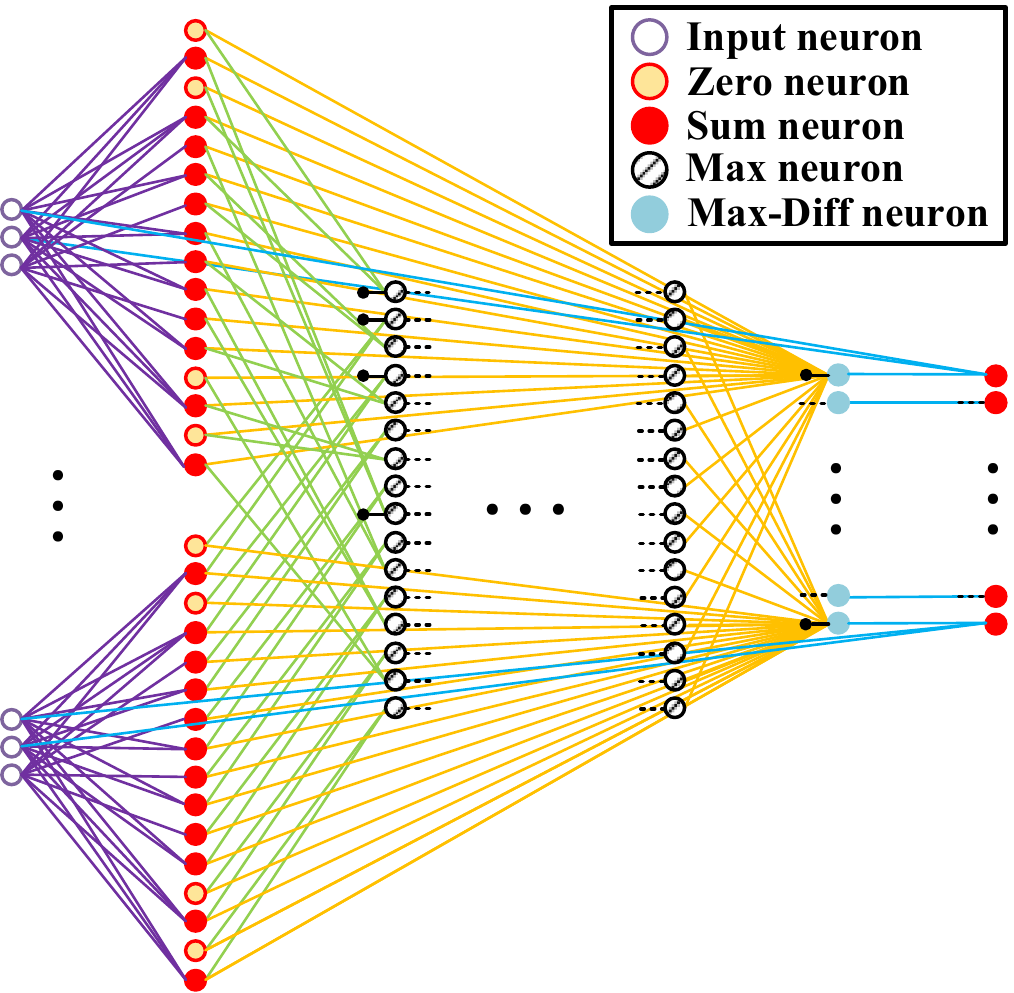}
	\caption{Subnet architecture based on the neural max-log-MAP algorithm. Some of the undrawn connecting lines are plotted as small dotted lines. Note that the neurons representing ${{{\bar{\alpha }}_{0}}({{s}'})}$ and ${{{\bar{\beta }}_{K}}(s)}$ are denoted as black circles.}
	\label{Fig:dnn decoder}
\end{figure}

The complete structure of the subnet is shown in Fig.~\ref{Fig:dnn decoder}. Given that the output of the \emph{M}th DNN decoding unit is ${{{L}^{M}}({{u}_{k}} |\mathbf{y})}$, the sigmoid function ${\sigma(x)\equiv{{( 1+{{e}^{-x}})}^{-1}}}$ is added, such that the final network output ${{{o}_{k}}=\sigma({{L}^{M}}({{u}_{k}} |\mathbf{y}))}$ is in the range of $[0,~1]$. 
Generally, the mean-squared error and binary cross-entropy can be used to calculate the network loss with ${{o}_{k}}$ and ${{u}_{k}}$. Nevertheless, the magnitude of the \emph{a posteriori} LLR calculated by the traditional max-log-MAP algorithm is usually greater than 10, whereas the sigmoid function is nearly close to 1 and 0 when $|x|>10$. Therefore, gradient vanishing is likely to occur if the loss is calculated with ${{o}_{k}}$. This situation will seriously affect the performance of the network and even prevent the network from converging. To overcome this problem, a redefined loss function computed as (\ref{Eq:loss}) is used to evaluate the loss of TurboNet
\begin{equation}
\label{Eq:loss}
\begin{aligned}
{\rm Loss}=\frac{1}{K}\sum\limits_{k=1}^{K}{{{\left( {{L}^{M}}\left( \left. {{u}_{k}} \right|\mathbf{y} \right)-{{{L}_{\rm log-MAP}^{T}} ({{u}_{k}} |\mathbf{y})} \right)}^{2}}},
\end{aligned}
\end{equation}
where ${{L}^{M}}({{u}_{k}}|\mathbf{y})$ represents the \emph{a posteriori} LLR obtained by the TurboNet consisting of \emph{M} decoding units and ${{{L}_{\rm log-MAP}^{T}} ({{u}_{k}} |\mathbf{y})}$ represents the \emph{a posteriori} LLR calculated by the traditional log-MAP algorithm with \emph{T} iterations. Notably, the iteration number \emph{T} of the target LLR ${{{L}_{\rm log-MAP}^{T}} ({{u}_{k}} |\mathbf{y})}$ can exceed the number of decoding units in the TurboNet and this feature can further help improve error-correction capabilities.

The aforementioned weights will be trained using the stochastic gradient descent (SGD) algorithm. The goal is to achieve minimal loss of the network by optimizing the trainable parameters ${\{w_{\bar{\gamma },({s}',s),k}^{i},w_{{s}',k}^{j},w_{({s}',s),k}^{j+1},w_{s,k}^{j+2},w_{{\rm e},k}^{l}\}}$, where $i=1,~2,~3$, $j=1,~4$, and $l=1,~2,~3$. The final decoding results can be obtained by a hard decision shown below:
\begin{equation}
\label{Eq:hard decision}
{{\hat{u}}_{k}}=\left\{ \begin{aligned}
& 1\text{\qquad }{{o}_{k}}\ge 0.5 \\ 
& 0\text{\qquad }{{o}_{k}}<0.5.
\end{aligned} \right.
\end{equation}

If setting all weights to 1, the results of~(\ref{Eq:weigh gamma}), (\ref{Eq:weigh posteriore llr}), and (\ref{Eq:weigh extrinsic llr}) will be the same as the original max-log-MAP algorithm. Hence, the performance of the TurboNet will not be inferior to the max-log-MAP algorithm and can improve significantly through training the network parameters. Moreover, the complexity of the TurboNet is similar to that of the turbo decoder using the max-log-MAP algorithm, especially when the code length is not long.

\section{Pruned DNN Decoder: TurboNet+}
\label{TurboNet+ design}
It is worth discussing to effectively reduce the number of parameters without sacrificing error-correction performance, that is, retain the most useful parameters as far as possible. Network pruning can potentially introduce two benefits to neural network compression and acceleration. Fewer parameters means less memory space and lower computing costs. In this section, we first statistically process the weights in (\ref{Eq:weigh gamma}), (\ref{Eq:weigh posteriore llr}), and (\ref{Eq:weigh extrinsic llr}). Then, a simplified weight pruning is applied to the TurboNet according to the weights histogram. We obtain a streamlined architecture called TurboNet+, which shows better convergence rate and bit-error rate (BER) performance. Finally, we study the overfitting issue through extensive simulation and present a training strategy to address the tricky issue.

\subsection{Weight Pruning}
The following conventions are employed to concisely distinguish different weights in TurboNet:
\begin{enumerate}
	\item [$\bullet$] GW: gamma weights defined in (\ref{Eq:weigh gamma})
	\item [$\bullet$] PLW: posterior LLR weights defined in (\ref{Eq:weigh posteriore llr})
	\item [$\bullet$] ELW: extrinsic LLR weights defined in (\ref{Eq:weigh extrinsic llr})
\end{enumerate}

\begin{figure}[t]
	\centering
	\includegraphics[width=4in]{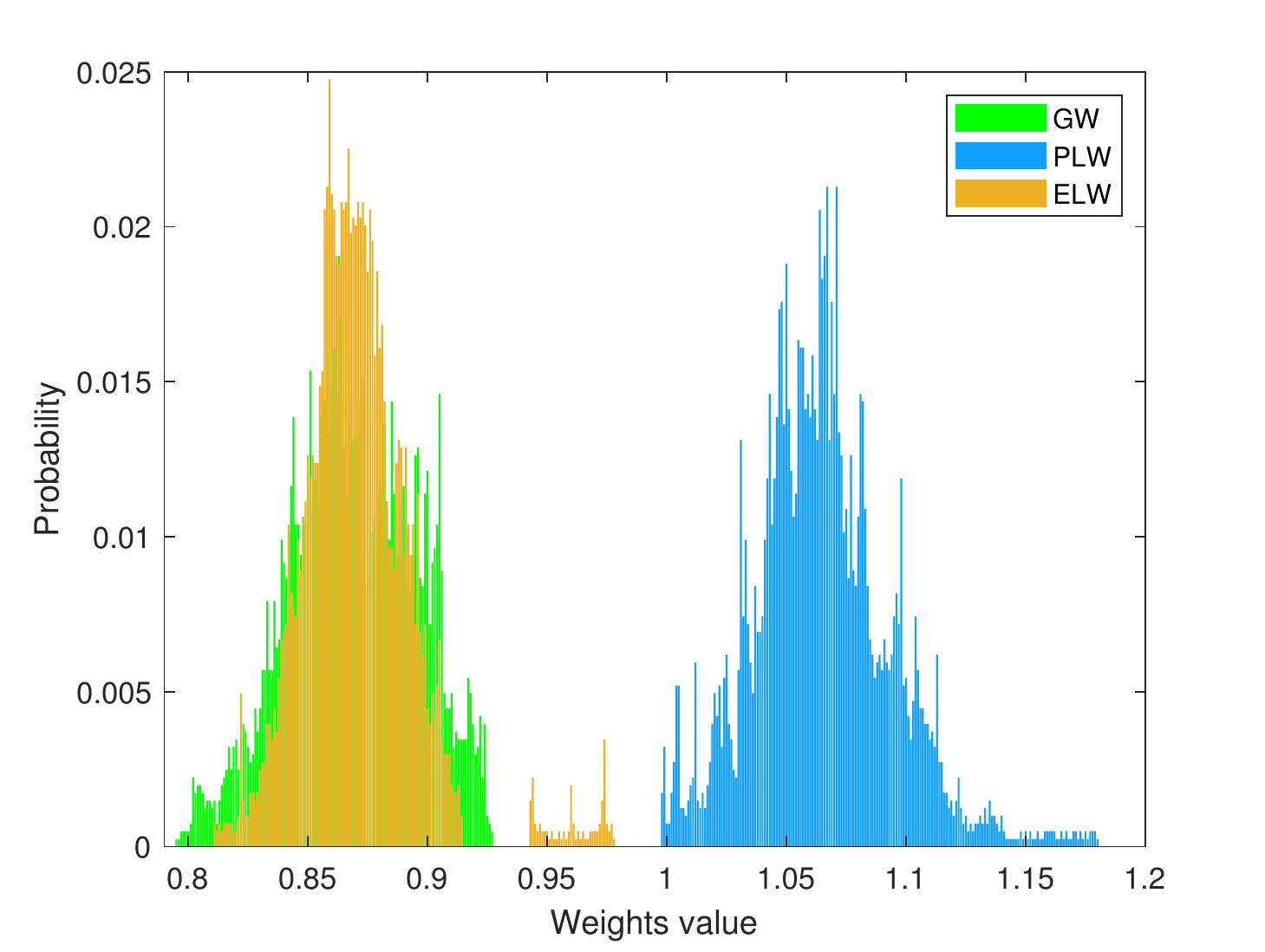}
	\caption{Weights histogram of DNN decoding unit 1 of the TurboNet for (40,~92) turbo code.}
	\label{Fig: weight histogram of TurboNet}
\end{figure}

Fig.~\ref{Fig: weight histogram of TurboNet} shows the weights histogram of the DNN decoding unit 1 of the trained TurboNet for (40,~92) turbo code\footnote{The code rate is ${1/2}$, and 12 bits are used to terminate the trellis}. Each column in the figure indicates the probability of the corresponding weights value. As shown in Fig.~\ref{Fig: weight histogram of TurboNet}, the weights assigned to (\ref{Eq:weigh gamma}), (\ref{Eq:weigh posteriore llr}), and (\ref{Eq:weigh extrinsic llr}) in the trained TurboNet are close to normal distribution. The trained TurboNet produces GW in the range from 0.8 to 0.92 while the max-log-MAP algorithm only has weights of size 1. The difference between the learned GW and initial values is obvious and the distribution is relatively uniform. Similarly, the ELW are distributed between 0.82 and 0.9 and mainly concentrate at 0.86. Compared with the GW, the distribution of the learned ELW is more concentrated. Different from the GW and the ELW, the PLW vary substantially between 1 and 1.1 and are mainly concentrated around 1.06. Thus, compared with the GW and the ELW, the difference between the PLW and initial values is not so significant. That is, the contribution of the PLW to the TurboNet is relatively small. The irregular activation function in (\ref{Eq:weigh posteriore llr}) may also have an unknown effect on the process of gradient descent. Therefore, discarding the PLW is rational. 

\begin{figure}[t]
	\centering
	\includegraphics[width=4in]{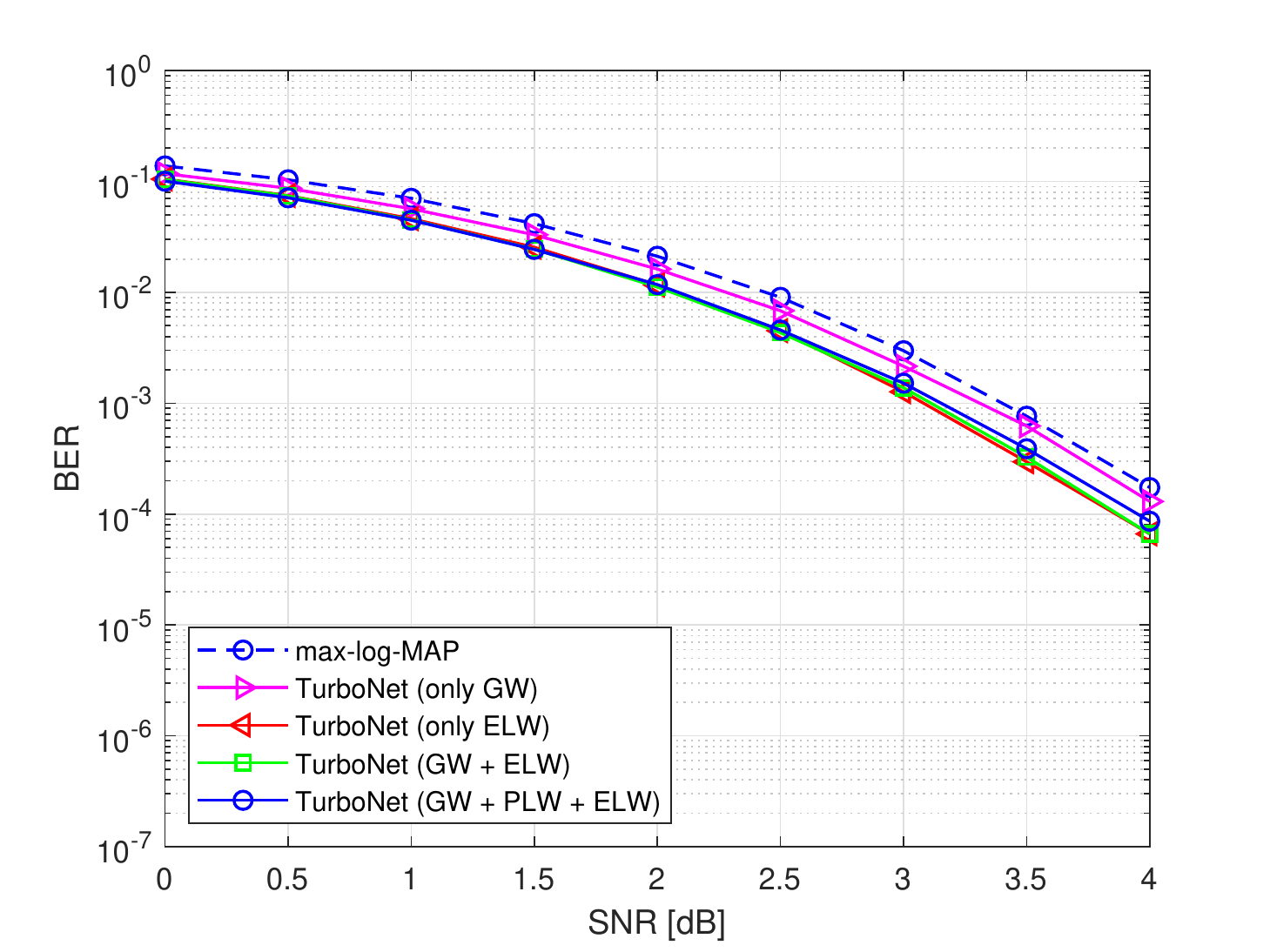}
	\caption{BER performance curve for (40,~92) turbo code using BPSK mapping on different scenarios.}
	\label{Fig:compare weights}
\end{figure}

In addition, we perform simulation to confirm this finding in four scenarios, i.e., (\textrm{i}) only retaining the GW, (\textrm{ii}) only retaining the ELW, (\textrm{iii}) retaining both the GW and the ELW, and (\textrm{iv}) using the GW, the PLW, and the ELW. The number of training samples is ${60,000}$ and the batch size is 500 for all scenarios. The numbers of training epochs for the scenarios (\textrm{i}), (\textrm{ii}), (\textrm{iii}), and (\textrm{iv}) are 10, 4, 5, and 50, respectively. The numerical results shown in Fig.~\ref{Fig:compare weights} further prove the significance of the ELW.

As depicted in Fig.~\ref{Fig:compare weights}, the TurboNet has achieved a certain gain compared with the max-log-MAP algorithm in all scenarios. The TurboNet discarding the PLW has obtained a similar gain compared with the one including the PLW at low SNRs. Moreover, the former shows better BER performance at high SNRs. Specifically, the TurboNet without the PLW achieves a gain of approximately 0.4~dB compared with the log-map-MAP algorithm with the same number of iterations, but achieves only approximately 0.3~dB with the PLW. Nevertheless, if we further abandon the ELW (i.e., only retain the GW), the improvement acquired by the TurboNet is not as obvious as before (only approximately 0.1~dB), regardless whether the SNR is low or high. Thus, the existing parameters are not sufficient to achieve satisfactory gains to compensate for the gap. Interestingly, if we further abandon the GW (i.e., only retain the ELW), the TurboNet cannot only acquire similar performance as the case retaining all parameters but can also obtain higher error-correction ability in certain cases. This improvement is most likely because the appropriate reduction in parameters makes the network easier to train. Overall, discarding the GW does not degrade the performance of the TurboNet, or alternatively, the ELW play a crucial role in performance improvement.

The above result coincides with~\cite{Improving max-log-MAP: Vogt}, where extrinsic information scaling is used to improve the decoding performance. From~\cite{Improving max-log-MAP: Vogt}, a scaling factor for the extrinsic LLRs, which is exchanged between the constituent SISO decoders, can improve the decoding quality. The reason for this improvement lies in the over-optimistic extrinsic LLR calculation of the max-log-MAP algorithm due to the omission of the logarithmic term.
In addition to the above advantages in the BER performance, the numbers of training epochs for different scenarios indicate that the convergence rate of the pruned network is significantly improved, thereby reducing the training overhead by an order of magnitude.

In summary, the above results demonstrate that the BER performance and convergence rate can be further improved by discarding the GW and the PLW. Therefore, we obtain an improved structure of the TurboNet called TurboNet+, which only retains the ELW. The subsequent discussions in this section are all for the TurboNet+ architecture.

\subsection{Training Strategy}
The SNR for training data affects the performance of the TurboNet+. One might use the same SNR for testing and training. But the accurate SNR is usually unavailable in reality.
To balance the performance and robustness, the SNR for training data usually varies within a certain range when training the network. However, this method suffers high complexity in actual implementation. As such, we should find low-complexity and sub-optimal SNR strategy with strong generalization. As in~\cite{IEEEondeep:Gruber}, the normalized validation error (NVE) can be used to determine the near optimal training SNR, defined as:
\begin{equation}
\label{Eq:NVE}
\begin{aligned}
{\rm NVE}\left( {{\rho }_{\rm train}} \right)=\frac{1}{L}\sum\limits_{l=1}^{L}{\frac{{\rm BER}_{{\rm DNN}}\left( {{\rho }_{\rm train}},{{\rho }_{{\rm test},l}} \right)}{{\rm BER}_{{\rm MAP}}\left( {{\rho }_{{\rm test},l}} \right)}},
\end{aligned}
\end{equation}
where ${{\rho }_{{\rm test},l}}$ represents the ${l\text{th}}$ SNR in a set of ${L}$ different validation data, ${{\rm BER}_{{\rm DNN}}({{\rho }_{\rm train}},{{\rho }_{{\rm test},l}})}$ denotes the BER obtained by a DNN trained at ${{\rho}_{\rm train}}$ on the data with ${{\rho}_{{\rm test},l}}$, and ${{\rm BER}_{{\rm MAP}}({{\rho }_{{\rm test},l}})}$ is the BER of MAP algorithm at ${{\rho}_{{\rm test},l}}$.
As indicated in~\cite{IEEEondeep:Gruber}, the NVE measures how good a DNN, trained at a particular SNR, is relative to the MAP algorithm over a range of different SNRs. 
However, the limitation of NVE is also clear, that is, we need extensive simulation to determine the SNR value.

It has been founded in~\cite{IEEEcommunicationalgorithms:Kim} that the training SNR obtaining the best BER for a target testing SNR also depends on the code rate. Furthermore, an experimental equation for the training of the neural decoders with binary phase shift keying (BPSK) has been obtained in~\cite{IEEEcommunicationalgorithms:Kim}, which can be expressed as
\begin{equation}
\label{Eq:min_choose_SNR}
\begin{aligned}
{{\rho }_{\rm train}}=\min \left\{ {{\rho }_{\rm test}},10{{\log }_{10}}\left( {{2}^{2R}}-1 \right) \right\},
\end{aligned}
\end{equation}
where ${R}$ denotes the code rate. Unfortunately, equation (\ref{Eq:min_choose_SNR}) is no longer applicable to higher order modulation, e.g., M-ary quadrature amplitude modulation (QAM). Therefore, the code rate is not the only factor to determine the optimal training SNR. Modulation orders, perhaps other factors, also affect it. Next, we study the influence of the training SNR and present an effective training strategy for the TurboNet+.

\begin{figure}[t]
	\centering
	\includegraphics[width=4in]{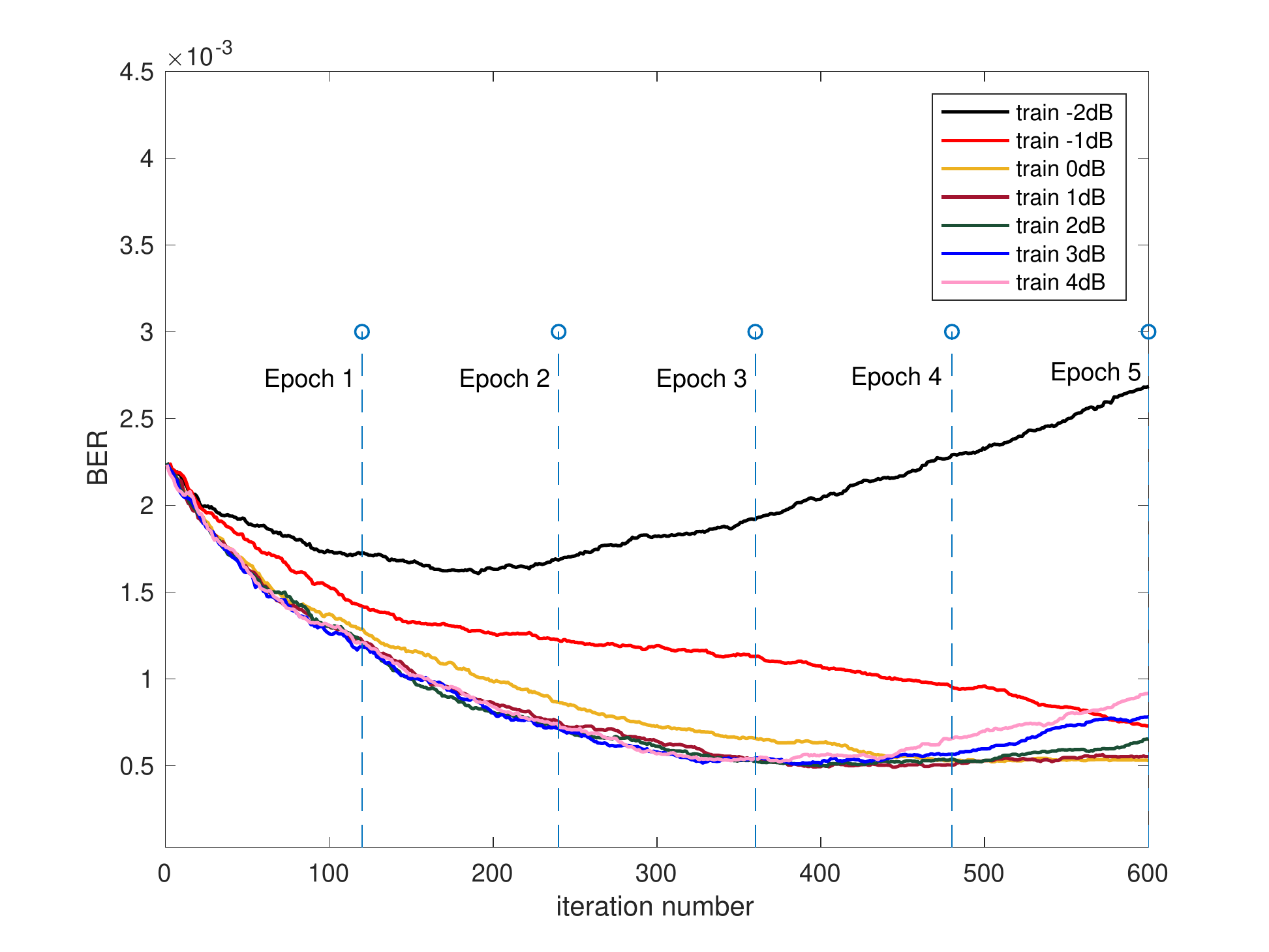}
	\caption{Learning curves for the TurboNet+ under different training SNRs.}
	\label{Fig: ber_vs_iteration_number_te3dB_R2}
\end{figure}

\begin{figure}[t]
	\centering
	\includegraphics[width=4in]{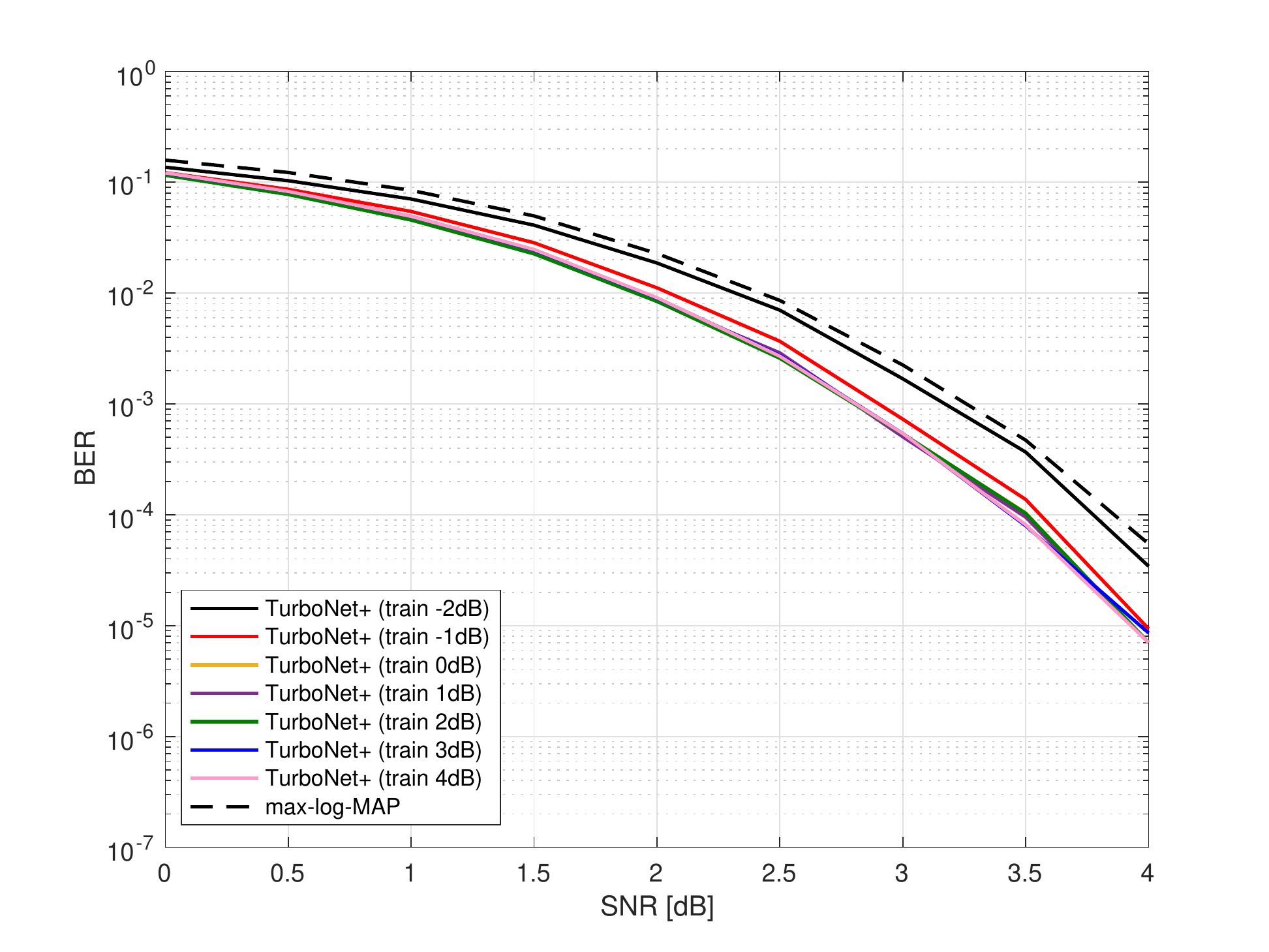}
	\caption{BER performance curves for the (64,~140) turbo code under different training SNRs. The training, validation, and testing sets contain $60,000$, $20,000$, and $200,000$ samples, respectively. The predefined maximum number of epochs is 5.}
	\label{Fig: compare_ber_with_different_tr_SNR_R2}
\end{figure}

We train the TurboNet+ for (64,~140) turbo code on randomly generated training data obtained over an AWGN channel, and ${{\rho} _ {\rm train}}$ ranges from --2~dB to 4~dB. The TurboNet+ consists of three DNN decoding units that correspond to three full iterations, and the loss function in~(\ref{Eq:loss}) is used with the target LLR ${{{L}_{\rm log-MAP}^{T}} ({{u}_{k}} |\mathbf{y})}$ being the log-MAP algorithm with $T=6$ iterations. The number of training data is ${{6}\times {{10}^{4}}}$ and the batch size is 500. The TurboNet+ is trained on 5 epochs with a learning rate of ${{8} \times {{10}^{-4}}}$ and the learning curves for the TurboNet+ with ${{{\rho }_{\rm test}}=3~\text{dB}}$ are shown in Fig.~\ref{Fig: ber_vs_iteration_number_te3dB_R2}. The figure indicates that when ${{\rho} _ {\rm train}=-2~\text{dB}}$, the TurboNet+ has not been well trained. Even though the TurboNet+'s convergence rate is relatively slow when ${{\rho}_{\rm train}}$ increases to $-1$~dB, it still converges to a near optimal performance. When ${{{\rho }_{\rm train}}=0~\text{dB}}$, the TurboNet+ can quickly converge to the optimal value and remain stable. The result obtained when ${{{\rho }_{\rm train}}=1~\text{dB}}$ is similar to the case when ${{{\rho }_{\rm train}}=0~\text{dB}}$. Notably, when ${{\rho}_{\rm train}=2,~3,~\text{and}~4~\text{dB}}$, the TurboNet+ still converges to the optimal value quickly, but overfits soon. Consequently, we use \emph{early stopping} to address the overfitting issue as follows:
\begin{enumerate}
	\item  Split the training data into a training set and a validation set, e.g. in a 3-to-1 proportion.
	\item  Train only on the training set and evaluate the BER performance on the validation set once in a while, e.g. after each epoch, and store the parameters if the performance is improved.
	\item Stop training as soon as the BER on the validation set increases or the predefined maximum number of epochs is met.
	\item Use the  latest stored parameters of the TurboNet+ as the result of the training.
\end{enumerate}
The results obtained according to the aforementioned training strategy are shown in Fig.~\ref{Fig: compare_ber_with_different_tr_SNR_R2}. The figure reveals that when ${{{\rho }_{\rm train}}=-2~\text{dB}}$, the improvement of the TurboNet+ is insignificant and when ${{{\rho }_{\rm train}}\ge-1~\text{dB}}$, the TurboNet+ has been well trained. Notably, when ${{\rho}_{\rm train}}$ ranges from 0~dB to 4~dB, the TurboNet+ obtains a similar performance.

Although the influence of the ${{\rho}_{\rm train}}$  on the convergence of the TurboNet+ can be ignored with the help of the proposed training strategy, we suggest that ${{\rho}_{\rm train}}$  should neither be too low nor too high. If ${{\rho}_{\rm train}}$ is too low, the TurboNet+ suffers poor performance because the network is very likely to be misled by the wrong information when the training data contain too many error samples. Conversely, if ${{\rho}_{\rm train}}$ is too high and the training does not stop in time, the overfitting issue may be extremely serious. We infer that those training samples that can be correctly decoded by the log-MAP algorithm with $T$ iterations but cannot by the max-log-MAP algorithm with $M$ iterations may be beneficial for improving the performance of the TurboNet+ because we use (\ref{Eq:loss}) as the loss function under supervised learning. If ${{\rho}_{\rm train}}$ is too high, most samples can be correctly decoded by both the max-log-MAP algorithm with $M$ iterations and the log-MAP algorithm with $T$ iterations. Consequently, the number of the training samples that meet the aforementioned requirement in the training set is negligible. We infer that this may cause the TurboNet+ easier to overfit when the training SNR is too high. Therefore, we can choose a moderate ${{\rho}_{\rm train}}$ and use \emph{early stopping} to stop training before overfitting.

\section{Simulation Results and Discussion}
\label{Simulation results and Discussion}
This section provides the simulation results of the TurboNet and the TurboNet+ under different code rates, code lengths, and modulation modes. The BER and complexity of the DL-based turbo decoding algorithms and traditional turbo decoding algorithms will be compared in detail.
\subsection{Parameter Settings}
The TurboNet and the TurboNet+ are constructed on top of the TensorFlow framework and an NVIDIA GeForce GTX 1080 Ti GPU is used for accelerated training. We train the TurboNet and the TurboNet+ for (40,~132) turbo code on randomly generated training data obtained over an AWGN channel at $0$~dB SNR with BPSK modulation. The TurboNet and the TurboNET+ consist of three DNN decoding units, corresponding to three full iterations. The loss function in~(\ref{Eq:loss}) is used and we set $T=6$ to obtain the target LLR ${{{L}_{\rm log-MAP}^{T}} ({{u}_{k}} |\mathbf{y})}$. We train the TurboNet and the TurboNet+ with SGD and the ADAM optimizer~\cite{IEEEadam:Kingma} with a batch size of 500. The learning rates of the TurboNet and the TurboNet+ are ${1 \times {10}^{-5}}$ and ${8 \times {10}^{-4}}$, respectively.

We also perform simulation on turbo codes with different code lengths at a higher code rate, i.e., ${R=1/2}$, and the training SNR for code lengths of 40 and 120 is 0.5~dB. In addition, we also verify the performance of the TurboNet and the TurboNet+ under higher-order modulation mode, i.e., 16-QAM, and the training SNR for code lengths of 40 and 120 is 6~dB.

\begin{figure}[t]
	\centering
	\includegraphics[width=4in]{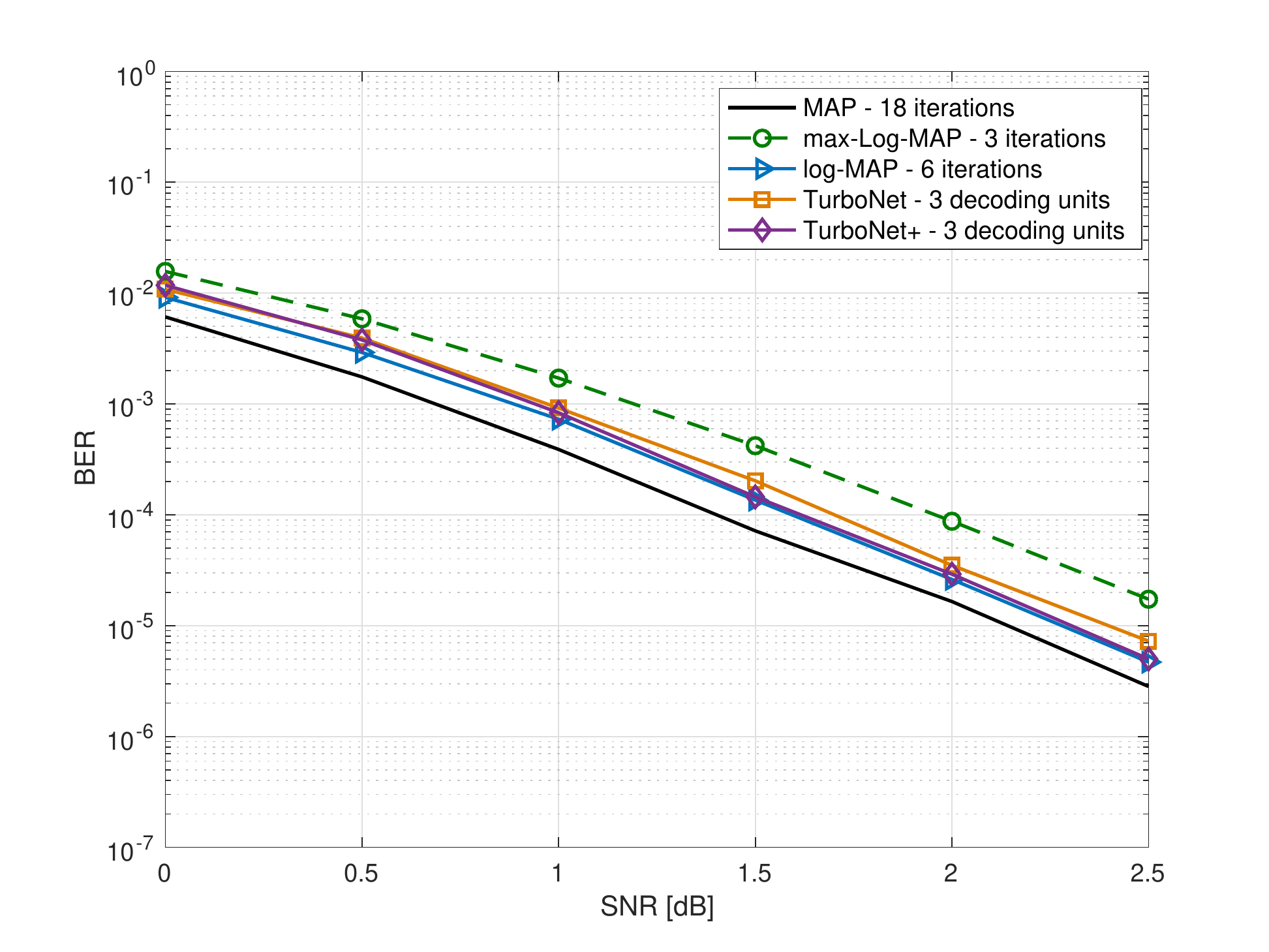}
	\caption{BER performance curves for (40,~132) turbo code with BPSK modulation. The training, validation, and testing sets contain $60,000$, $20,000$, and $2,000,000$ samples, respectively. The maximum numbers of training epochs for TurboNet and TurboNet+ are 50 and 10, respectively.}
	\label{Fig: BER of TurboNet/TurboNet+ BPSK R_1/3}
\end{figure}

\subsection{BER Performance}
The BER performance curves for the (40,~132) turbo code using different decoding algorithms are shown in Fig.~\ref{Fig: BER of TurboNet/TurboNet+ BPSK R_1/3}.
From the figure, the MAP algorithm with 18 iterations obtains the best BER performance  among all turbo decoding algorithms, which outperforms the max-log-MAP algorithm with three iterations by approximately 0.6~dB. The log-MAP algorithm with six iterations achieves the sub-optimal performance. Next are the TurboNet+ with three decoding units and the TurboNet with three decoding units, which are both supervised by the log-MAP algorithm with six iterations. The TurboNet+ with three decoding units outperforms the max-log-MAP algorithm with the same number of iterations at all SNR ranges and approximates the log-MAP algorithm with six iterations. In addition, Fig.~\ref{Fig: BER of TurboNet/TurboNet+ BPSK R_1/3} shows that the gap between the TurboNet+ decoder and the traditional max-log-MAP algorithm will gradually increase with the SNR, that is, reaching 0.25~dB at low SNRs and 0.4~dB at high SNRs. The result also indicates that the TurboNet+ could reduce the complexity of the TurboNet without sacrificing the BER performance, which is consistent with the conclusion obtained in Section~\ref{TurboNet+ design}.
\begin{figure}[t]
	\centering
	\includegraphics[width=4in]{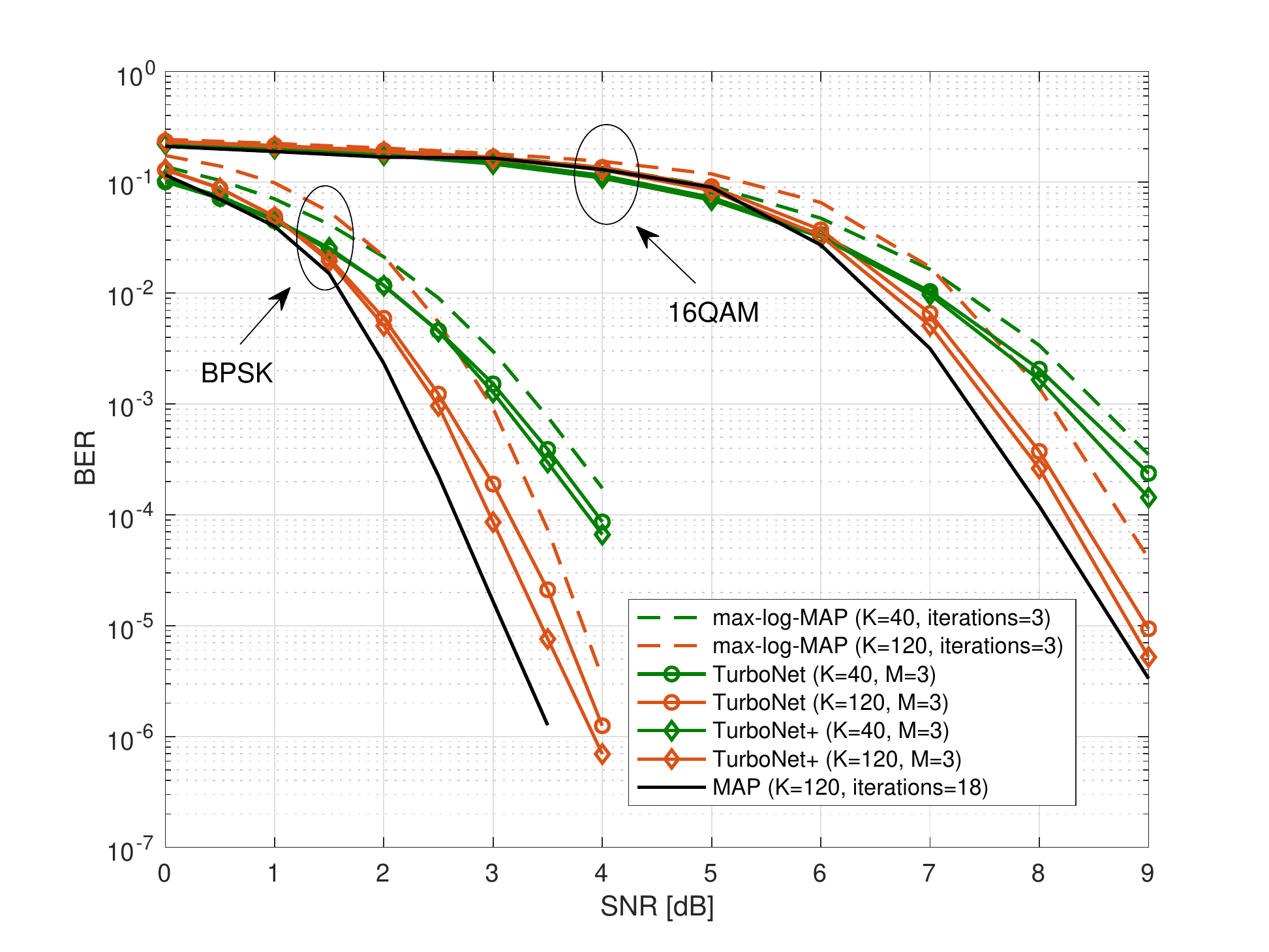}
	\caption{BER performance curves for turbo codes with different information bit lengths and modulation modes. The training, validation, and testing sets contain $60,000$, $20,000$, and $2,000,000$ samples, respectively. The maximum numbers of training epochs for TurboNet and TurboNet+ are 50 and 10, respectively.}
	\label{Fig:ber of TurboNet+}
\end{figure}
Fig.~\ref{Fig:ber of TurboNet+} compares the BER performance of the DL-based turbo decoders and the traditional max-log-MAP algorithm under different code lengths and modulation modes, where the code rate ${R=1/2}$ and the MAP algorithm with 18 iterations is plotted as the benchmark.
The TurboNet+ continues to achieve the best BER performance compared with the TurboNet decoder and the traditional max-log-MAP algorithm with the same number of iterations, and such outcome is similar to the situation when the code rate $R=1/3$. Fig.~\ref{Fig:ber of TurboNet+} suggests that even though the TurboNet+ with three decoding units is trained at a single SNR, it can outperform the max-log-MAP algorithm and the TurboNet with the same number of iterations in a wide SNR range. As the code length increases, the advantages of the TurboNet+ over the TurboNet in BER performance are also more obvious. Specifically, as the information bit length, $K$, increases, the improvement achieved by the TurboNet+ will be more significant, i.e., 0.4~dB for $K=40$ and 0.5~dB for $K=120$. We speculate that this result arises because the error-correction capability of the max-log-MAP algorithm itself is limited when the information bit length is short and the improvement is not substantial. As the information bit length increases, the error-correction capability of the max-log-MAP algorithm is enhanced, and the improvement of the TurboNet+ is more obvious. Fig.~\ref{Fig:ber of TurboNet+} also reveals that when the modulation order increases, the improvement of the TurboNet+ at low SNRs is not as significant as that of BPSK. However, at high SNRs, the TurboNet+ can still achieve distinct gains compared with the max-log-MAP algorithm and reaches 0.5~dB for $K=120$.
Similarly, the TurboNet+'s gain becomes more pronounced as the information bit length increases.

\begin{figure}[t]
	\centering
	\includegraphics[width=4in]{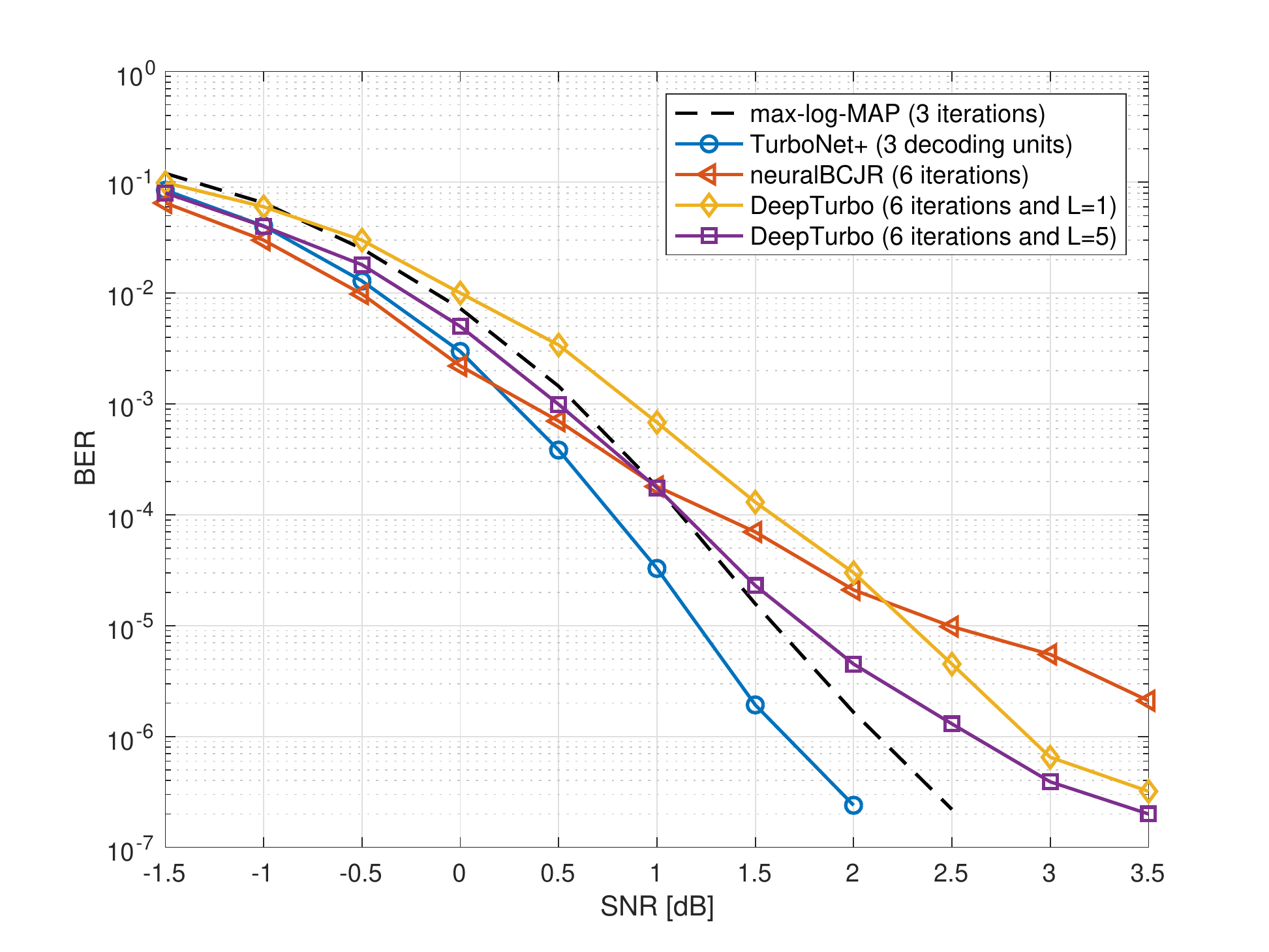}
	\caption{BER performance curves of different turbo decoders. $L$ denotes the information passed to the next stage for one bit position~\cite{Jiang-DeepTurbo}.}
	\label{Fig: BER performance curves of different turbo decoders.}
\end{figure}

Fig.~\ref{Fig: BER performance curves of different turbo decoders.} compares the BER performance of the model-driven turbo decoders and the data-driven turbo decoders, where the block length is 100 and the code rate is $1/3$. The SISO decoders in the neuralBCJR~\cite{IEEEcommunicationalgorithms:Kim} and DeepTurbo~\cite{Jiang-DeepTurbo} decoders are two bidirectional long short-term memory (LSTM) layers, and the number of hidden units in each LSTM layer is 800. The result suggests that the performance of different DL-based turbo decoders is relatively close in low SNRs. The neuralBCJR algorithm with six iterations obtains the best BER performance and the TurboNet+ decoder with three decoding units has achieved the sub-optimal performance. With the increase of the SNR, the advantages of the model-driven turbo decoder over the data-driven turbo decoders becomes apparent gradually. Notably, the TurboNet+ with three decoding units outperforms the optimal data-driven decoder by approximately 1.5~dB at high SNRs. In general, the gap between the model-driven and data-driven approaches is significant, which benefits from the reasonable integration of the TurboNet+ with the max-log-MAP algorithm and DL tools.

\subsection{Computational Complexity}
For the SN1 in the TurboNet, the number of weights introduced in the hidden layer 1, hidden layer $K+1$, and output layer is $24K$, $48K$, and $3K$, respectively. Therefore, a subnet in the TurboNet contains a total of $75K$ weights, and the TurboNet with $M$ decoding units contains $ 2M \times75K = 150MK $ weights (a decoding unit contains two subnets). In addition, the pruned TurboNet, i.e., TurboNet+, contains $3K$ weights per subnet. We can obtain that the TurboNet+ with $M$ decoding units contains a total of $ 2M \times3K = 6MK$ weights. Consequently, TurboNet+ has a small number of weights, which indicates that we can store some weights for different block lengths in advance to meet practical requirements.

\begin{table}[b]
	\caption{Complexity analysis for turbo decoders }
	\label{tab:Computational Complexity}
	\centering
\begin{tabular}{cccc}
	\toprule
	decoding algorithm & \# of iterations  & \# of parameters & time (seconds) \\
	\hline
	max-log-MAP & 3 & - & $3.02\times {10}^{-3}$ \\
	log-MAP & 6 & - & $5.93\times {10}^{-2}$ \\
	TurboNet & 3 & $4.5\times {10}^4$ & $3.22\times {10}^{-3}$ \\
	TurboNet+ & 3 & $1.8\times {10}^3$ & $3.04\times {10}^{-3}$ \\
	neuralBCJR in~\cite{IEEEcommunicationalgorithms:Kim} & 6 & $7.8 \times {10}^6$ & 1.67\\
	DeepTurbo ($L=5$) in~\cite{Jiang-DeepTurbo} & 6  & $9.4\times {10}^7$ & 1.84\\ 
	\bottomrule
\end{tabular}
\end{table}

In Table~\ref{tab:Computational Complexity}, we compare the complexities of different turbo decoders in terms of the overall time consumption required to complete a single-forward pass of one codeword. Time comparison is made on a computer with OSX 10.12, i5-6360U 2.9GHz dual-core CPU and 8 GB RAM. We set $K=100$ and $R=1/3$. As shown in Table~\ref{tab:Computational Complexity}, the computational complexity is increased due to TurboNet's blind introduction of parameters and this feature may become more distinct when $K$ increases, although the TurboNet can improve the BER performance of the max-log-MAP algorithm. However, the TurboNet+ significantly improves the BER performance of the max-log-MAP algorithm and the introduced computational complexity is negligible. Specifically, the TurboNet+ with three decoding units can approximate the BER performance of the log-MAP algorithm with six iterations, but perform nearly 20 times faster than the log-MAP algorithm with six iterations. Thus,  the TurboNet+ can reduce decoding latency significantly,  which is of great significance for low-latency communications. Notably, the TurboNet+ benefits from the fact that its parameters are negligible and has a much lower computational cost compared with the data-driven neuralBCJR decoder~\cite{IEEEcommunicationalgorithms:Kim} and DeepTurbo~\cite{Jiang-DeepTurbo}. Given that the TurboNet+ has achieved the best BER performance with low computational overhead in all simulation scenarios, this structure will be used in the subsequent OTA test.

\section{OTA Test and Result Discussion}
\label{OTA Test}
From the above simulation results, the TurboNet+ offers higher error-correction capability than the traditional max-log-MAP algorithm. Another aspect of the proposed algorithm is its performance in real environments.

There have been several prototyping systems to verify the effectiveness and robustness\cite{AI OFDM Receiver: Jiang, AI receiver: Zhang}. In~\cite{RaPro: Yang}, a novel fifth-generation rapid prototyping~(RaPro) system architecture has been suggested to deploy FPGA-privileged modules on software-defined radio (SDR) platforms and has proven to be highly flexible and scalable. In this section, we use the RaPro system as our testbed to evaluate the BER performance of our TurboNet+ in real channel environments.
\subsection{System Setup}
\begin{figure}[t]
	\centering
	\includegraphics[width=5.7in]{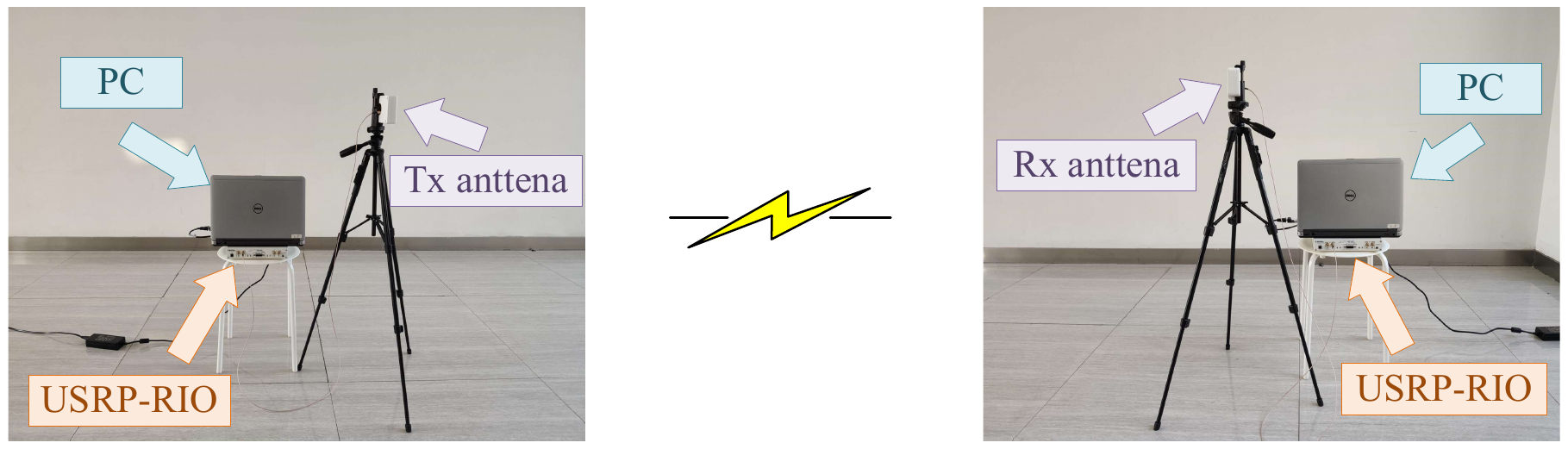}
	\caption{Real testing system consisting of transmitter and receiver.}
	\label{Fig: Real_Testing_Scenario}
\end{figure}

As shown in Fig.~\ref{Fig: Real_Testing_Scenario}, the real testing system contains a transmitter and a receiver, which offers transmission and reception of radio frames composed of OFDM symbols. We employ two SDR nodes of the universal software radio peripheral reconfigurable I/O (USRP-RIO) series manufactured by National Instruments. Each SDR node includes one RF transceiver of 40 MHz bandwidth and a programmable FPGA responsible for distributed signal processing, such as the reciprocity calibration or OFDM (de)modulation~\cite{RaPro: Yang}. We perform the OTA indoor test. The details of the test scenarios are discussed in next subsection. 

\begin{figure}[t]
	\centering
	\includegraphics[width=4.5in]{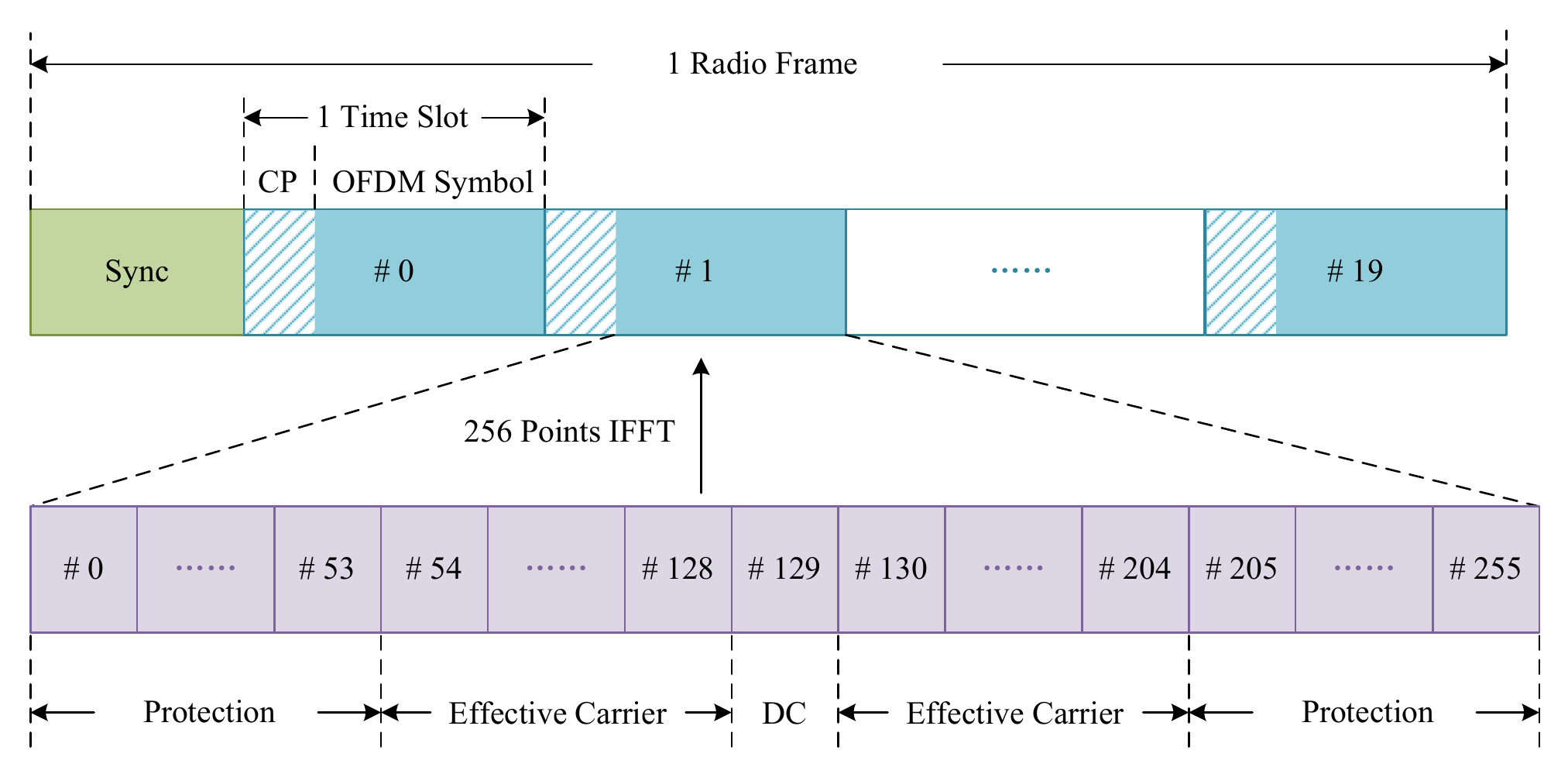}
	\caption{Frame structure for transmitting data.}
	\label{Fig: Frame_Structure}
\end{figure}

Fig.~\ref{Fig: Frame_Structure} illustrates the radio frame structure at the transmitter. One radio frame contains 20 time slots and a frame head. Each time slot corresponds to one OFDM symbol and its corresponding cyclic prefix~(CP) and the frame head is added for synchronization. The lengths of each OFDM symbol and CP are 256 and 64, respectively. Each OFDM symbol is obtained by performing 256-point inverse fast Fourier transform (IFFT) on the OFDM symbol in the frequency domain. In the frequency domain, each OFDM symbol  contains 256 subcarriers, among which 150 subcarriers are effective for transmitting pilot symbols and data symbols, 105 subcarriers are employed as the guard band, and one subcarrier is the direct current (DC) offset. The 150 effective subcarriers consist of 25 pilot symbols and 125 data symbols. Comb type pilots are used to insert pilot symbols and each pilot symbol is followed by 5 data symbols.

Since one radio frame contains 20 time slots, each with 125 data symbols, we can transmit $20\times 125=2,500$ data symbols in each radio frame. For simplicity, the modulation mode of the quadrature-phase-shift keying (QPSK) is applied in the real testing system. Thus, a total of $2\times2,500=5,000$ bits, called transmission bits, are transmitted per radio frame. Those transmission bits consist of two frame number blocks, 40 data blocks, and one zero block. We denote 16 bits as a frame number so that the receiver knows which frame is received. Two 16-bit frame numbers and 8 bits of 0 constitute 40-bit information bits, which are passed through a turbo encoder with a code rate of $1/2$ to obtain a 92-bit codeword. A frame number block with a length of 100 bits is obtained by zero padding to the codeword. Each data block of 100 bits is similarly obtained from randomly generated information bits. Specifically, we generate information bits of length 40 randomly and the information bits are passed through a turbo encoder with a code rate of $1/2$ to obtain a 92-bit codeword. A data block with a length of 100 bits is obtained by zero padding to the codeword similarly. The zero block contains 800 bits of 0 for noise estimation.

At the transmitter side, the transmitting bits are QPSK-modulated and IFFT-based OFDM modulated to obtain OFDM symbols in the time domain. After the CP is added to the corresponding OFDM symbol, the radio frame is transmitted by an USRP-RIO through a RF antenna whose center frequency is set to 2~GHz. At the receiver side, wireless signals are received by an USRP-RIO through a RF antenna.  The receiver first conducts frame detection. After CP removal and FFT-based OFDM demodulation, the USRP-RIO performs least-square channel estimation. Finally, the results are stored into .csv files for future use.

\subsection{Experimental Results}
The entire OTA test is divided into two phases. The first phase obtains samples in different scenarios with the help of the RaPro system shown in Fig.~\ref{Fig: Real_Testing_Scenario}. The second phase employs the network trained under different training sets to verify the robustness of the TurboNet+ in real channels. We chose four different scenarios to evaluate our TurboNet+ in real environments.
\begin{enumerate}
	\item[$\bullet$] Scenario~1: fixed indoor scenario. The receiver is 4 meters away from the transmitter in a room with windows, walls, and corridor around;
	\item[$\bullet$] Scenario~2: an indoor scenario. Unlike Scenario~1, the antennas are 5 meters apart and are not facing each other;
	\item[$\bullet$] Scenario~3: a changing indoor scenario. Unlike Scenario~2, pedestrians keep walking between the antennas;
	\item[$\bullet$] Scenario~4: a changing indoor scenario. The only difference from Scenario~3 is that the antennas are 6 meters apart.
\end{enumerate}

The measured BER is shown in Table~\ref{tab:BER Performances2}. From the Table~\ref{tab:BER Performances2}, the models trained in matched scenarios always achieve the best performance, thereby reflecting the TurboNet+’s strong ability to learn new scenarios. Moreover, the training samples obtained at a single transmitting power can help the TurboNet+ to achieve significant improvement over a wide range of transmitting power, a feature that is consistent with the simulation results. Table~\ref{tab:BER Performances2} shows that the model trained in Scenario~1 can achieve similar performance in new scenarios compared with the retrained model and this outcome indicates the TurboNet+’s strong robustness. Note that the model trained with samples generated by simulation under an AWGN channel can also achieve decent results in the real scenarios. Although the gain obtained cannot meet the above two models, a considerable improvement is achieved compared with the traditional max-log-MAP algorithm. In general, the TurboNet+ exhibits strong learning ability and robustness under real channels.

\begin{table}[t]
	\caption{BER performance of turbo decoders for OTA test}
	\label{tab:BER Performances2}
	\centering
	\begin{threeparttable}
				\begin{tabular}{ccccccccc}
			\toprule
			& & -1~dBm & 0~dBm & 1~dBm & 2~dBm & 3~dBm & 4~dBm & 5~dBm\\
			\hline
			\multirow{3}{*}{\rotatebox{0}{Scenario 1}} & max-log-MAP & 3.0e-2 & 1.4e-2 & 9.1e-3 & 6.0e-3 & 4.3e-3 & 2.6e-4 & 4.6e-5\\
			& TurboNet+\tnote{1} & 1.9e-2 & 8.3e-3 & 4.8e-3 & 3.2e-3 & 2.3e-3 & 1.3e-4 & 2.7e-5\\
			& TurboNet+\tnote{2} & 1.6e-2 & 6.5e-3 & 3.7e-3 & 2.4e-3 & 1.7e-3 & 1.1e-4 & 2.1e-5\\
			\hline
			\multirow{4}{*}{\rotatebox{0}{Scenario 2}} & max-log-MAP & 2.8e-2 & 1.4e-2 & 6.6e-3 & 2.5e-3 & 1.2e-3 & 3.4e-4 & 9.8e-5\\
			& TurboNet+\tnote{1} & 1.9e-2 & 8.9e-3 & 4.2e-3 & 1.5e-3 & 8.1e-4 & 2.4e-4 & 6.4e-5\\
			& TurboNet+\tnote{2} & 1.8e-2 & 8.8e-3 & 4.0e-3 & 1.5e-3 & 7.6e-3 & 2.3e-4 & 6.2e-5\\
			& TurboNet+\tnote{3} & 1.7e-2 & 8.0e-3 & 3.9e-3 & 1.4e-3 & 7.5e-4 & 2.2e-4 & 5.8e-5\\
			\hline
			\multirow{4}{*}{\rotatebox{0}{Scenario 3}} & max-log-MAP & 2.3e-2 & 1.5e-2 & 5.0e-3 & 3.5e-3 & 7.0e-4 & 3.5e-4 & 2.3e-4\\
			& TurboNet+\tnote{1} & 1.6e-2 & 9.8e-3 & 3.2e-3 & 2.3e-3 & 4.6e-4 & 2.3e-4 & 1.5e-4\\
			& TurboNet+\tnote{2} & 1.5e-2 & 9.5e-3 & 3.1e-3 & 2.2e-3 & 4.2e-4 & 2.2e-4 & 1.5e-4\\
			& TurboNet+\tnote{3} & 1.4e-2 & 8.9e-3 & 2.8e-3 & 2.1e-3 & 4.2e-4 & 2.1e-4 & 1.4e-4\\
			\hline
			\multirow{4}{*}{\rotatebox{0}{Scenario 4}} & max-log-MAP & 3.6e-2 & 1.9e-2 & 9.0e-3 & 6.7e-3 & 3.7e-3 & 1.3e-3 & 6.8e-4\\
			& TurboNet+\tnote{1} & 2.6e-2 & 1.3e-2 & 6.0e-3 & 4.4e-3 & 2.3e-3 & 8.7e-4 & 4.4e-4\\
			& TurboNet+\tnote{2} & 2.6e-2 & 1.2e-3 & 5.9e-3 & 4.2e-3 & 2.1e-3 & 8.3e-4 & 4.3e-4\\
			& TurboNet+\tnote{3} & 2.5e-2 & 1.2e-2 & 5.5e-3 & 4.0e-3 & 2.1e-3 & 8.0e-4 & 4.2e-4\\
			\bottomrule
		\end{tabular}
		\begin{tablenotes}
			\item[1] The model trained with the data generated by MATLAB simulation under an AWGN channel.
			\item[2] The model trained with the data collected by OTA test in Scenario~1.
			\item[3] The model trained with the data collected by OTA test in corresponding scenario. 
		\end{tablenotes}
	\end{threeparttable}
\end{table}

\section{Conclusions}
\label{Conslusion}
In this work, we have proposed the TurboNet, which was obtained by unfolding the original iterative structure of the max-log-MAP algorithm as a DNN decoding unit. In particular, we introduced weights to the max-log-MAP algorithm and trained the weights through a supervised learning algorithm. A loss function was well-designed to efficiently train the TurboNet and prevent tricky gradient vanishing issue. We further reduced the computational complexity and training cost by pruning the TurboNet into the TurboNet+. The TurboNet+ has significant advantages in computational complexity compared with the existing RNN decoder, and such a feature is conducive to significantly reducing decoding overhead. We also provided a simple and effective training strategy that could enable the proposed TurboNet+ to be trained better and faster. Simulation results demonstrated TurboNet+’s superiority in error-correction ability and computational overhead. The OTA test further proved TurboNet+’s strong robustness.

\ifCLASSOPTIONcaptionsoff
\newpage
\fi

\end{document}